\documentclass[10pt,twocolumn,showpacs,preprintnumbers,amsmath,amssymb,aps,prl,superscriptaddress,longbibliography]{revtex4-2}
\usepackage{appendix}
\usepackage{bbm}
\usepackage{mathrsfs}
\usepackage{graphicx}
\usepackage{dcolumn}
\usepackage{bm}
\usepackage{braket}
\usepackage{amsmath}
\usepackage{amsfonts}
\usepackage[dvipsnames]{xcolor}
\usepackage[colorlinks=true,linkcolor=Blue,urlcolor=BlueViolet,citecolor=BlueViolet]{hyperref}
\usepackage{natbib}
\usepackage{multirow}

\begin{document}

\title{Protect Measurement-Induced Phase Transition from Noise}
\author{Dongheng Qian}
\affiliation{State Key Laboratory of Surface Physics and Department of Physics, Fudan University, Shanghai 200433, China}
\affiliation{Shanghai Research Center for Quantum Sciences, Shanghai 201315, China}
\author{Jing Wang}
\thanks{Contact author: wjingphys@fudan.edu.cn}
\affiliation{State Key Laboratory of Surface Physics and Department of Physics, Fudan University, Shanghai 200433, China}
\affiliation{Shanghai Research Center for Quantum Sciences, Shanghai 201315, China}
\affiliation{Institute for Nanoelectronic Devices and Quantum Computing, Fudan University, Shanghai 200433, China}
\affiliation{Hefei National Laboratory, Hefei 230088, China}

\begin{abstract}
Scrambling dynamics induced by random unitary gates can protect information from low-rate measurements, which underpins the phenomenon known as the measurement-induced phase transition (MIPT). However, typical decoherence noises disrupts the volume law phase, complicating the observation of MIPT on noisy intermediate-scale quantum devices. Here, we demonstrate that incorporating quantum-enhanced operations can effectively protect MIPT from environmental noise, thereby enabling its detection in experiment. The transition is characterized by the conditional entanglement entropy (CEE), which is associated with a statistical mechanics model wherein noise and quantum-enhanced operations act as competing external random fields. When the net external field is zero, a ferromagnetic-paramagnetic phase transition is expected, resulting in the MIPT. This zero-field condition also ensures an \emph{average} apparatus-environment symmetry, making CEE a valid probe of entanglement and establishing the transition as a genuine entanglement phase transition. Additionally, we provide numerical results demonstrate the MIPT in a (2+1)-dimensional quantum circuit under dephasing noise. We also propose a method to estimate the noise rate, enabling the zero-field condition to be achieved experimentally and ensuring the feasibility of our protocol. Our result serves as a concrete example of the power of quantum enhancement in combating noise.

\end{abstract}

\date{\today}

\maketitle

Noise presents one of the most significant threats to reliable quantum computation and the long-term storage of quantum information. To address this challenge, quantum error correction (QEC) techniques are employed~\cite{schumacher1996, kitaev1997, kitaev2003a}. The core idea of QEC is to encode quantum information in a noise-resilient manner, allowing errors induced by noise to be effectively detected and corrected. The encoding schemes and correction procedures in QEC are typically highly structured~\cite{dennis2002, wang2003}.

In recent years, measurement-induced phase transitions (MIPT) have attracted significant attention~\cite{li2019, gullans2020a, chan2019, bao2024, bao2021b, garratt2023b, lee2022, li2021a, li2023a, nahum2020, nahum2018, nahum2019, sharma2022, sang2021a, skinner2019, szyniszewski2019, vasseur2019, zabalo2020, alberton2021, fidkowski2021, fisher2023, poboiko2024, qian2024, yu2022}. MIPTs can be understood from the perspective of QEC, where information is encoded in an unstructured way through scrambling by random unitary gates, with measurements identified as sources of noise~\cite{choi2020}. When the measurement rate is low, corresponding to a low error rate, the entanglement entropy of the state follows a volume law, indicating that the information remains protected. Although this transition has been demonstrated in several experiments~\cite{hoke2023, kamakari2024, koh2023, noel2022}, a major obstacle to observing this transition is its instability against various decoherence noises. It has been shown that the state obeys an area law even with an infinitesimal rate of dephasing noise or resetting noise in the circuit's bulk~\cite{dias2023, liu2023e, weinstein2022, liu2024}. This can be interpreted as the scrambling dynamics being insufficient to protect quantum information against these more common practical noises. Therefore, it is both experimentally relevant and theoretically interesting to find a way to protect MIPT from decoherence noise.

Meanwhile, quantum-enhanced (QE) operation has emerged as a potentially more powerful and flexible method for extracting information about a quantum state compared to traditional projective measurements, which only allow access to classical information~\cite{aharonov2022, huang2021b, zhao2017, braun2018}. The fundamental idea of quantum enhancement is to use a quantum sensor, rather than a classical sensor, to detect the system. It has been demonstrated that an exponential speedup can be achieved in certain tasks by coherently manipulating these quantum probes~\cite{aharonov2022}. However, it remains an open question whether these QE operations and algorithms can maintain their advantage in the presence of noise.

In this Letter, we demonstrate that the original MIPT can be protected against decoherence caused by environmental noise through QE operations. From the perspective of QEC, these QE operations act as an unstructured protocol to protect quantum information by encoding it within both the system qubits and the ancilla qubits. 

We first provide analytical analysis, mapping the conditional entanglement entropy (CEE) to the free energy of a statistical-mechanics model, where noise and QE operation exactly correspond to two symmetry-breaking fields in different directions. Combining with unitary gate and measurement, we obtain a random field model with ferromagnetic coupling. 
This model exhibits a ferromagnetic-paramagnetic phase transition driven by both increasing temperature and increasing random field strength, under the condition of zero net external field. This phase transition thus leads to the transition in the scaling of CEE. Next, we discover that the zero-field condition is equivalent to an \emph{average} apparatus-environment exchange (aAEE) symmetry, which ensures the CEE to be a valid probe of entanglement~\cite{kelly2024a, kelly2024b}. Consequently, the transition observed in CEE directly corresponds to an entanglement phase transition. By imposing the aAEE symmetry, we conduct numerical studies for (2+1)-d circuits and explicitly demonstrate the MIPT in the presence of various decoherence noises. Our approach is inspired by Ref.~\cite{kelly2024a}, where a prototype setting was introduced. The structure we introduced here incorporates measurement and is much easier to implement in practice, as will be further elaborated later.

\begin{figure}[t]
\begin{center}
\includegraphics[width=3.4in, clip=true]{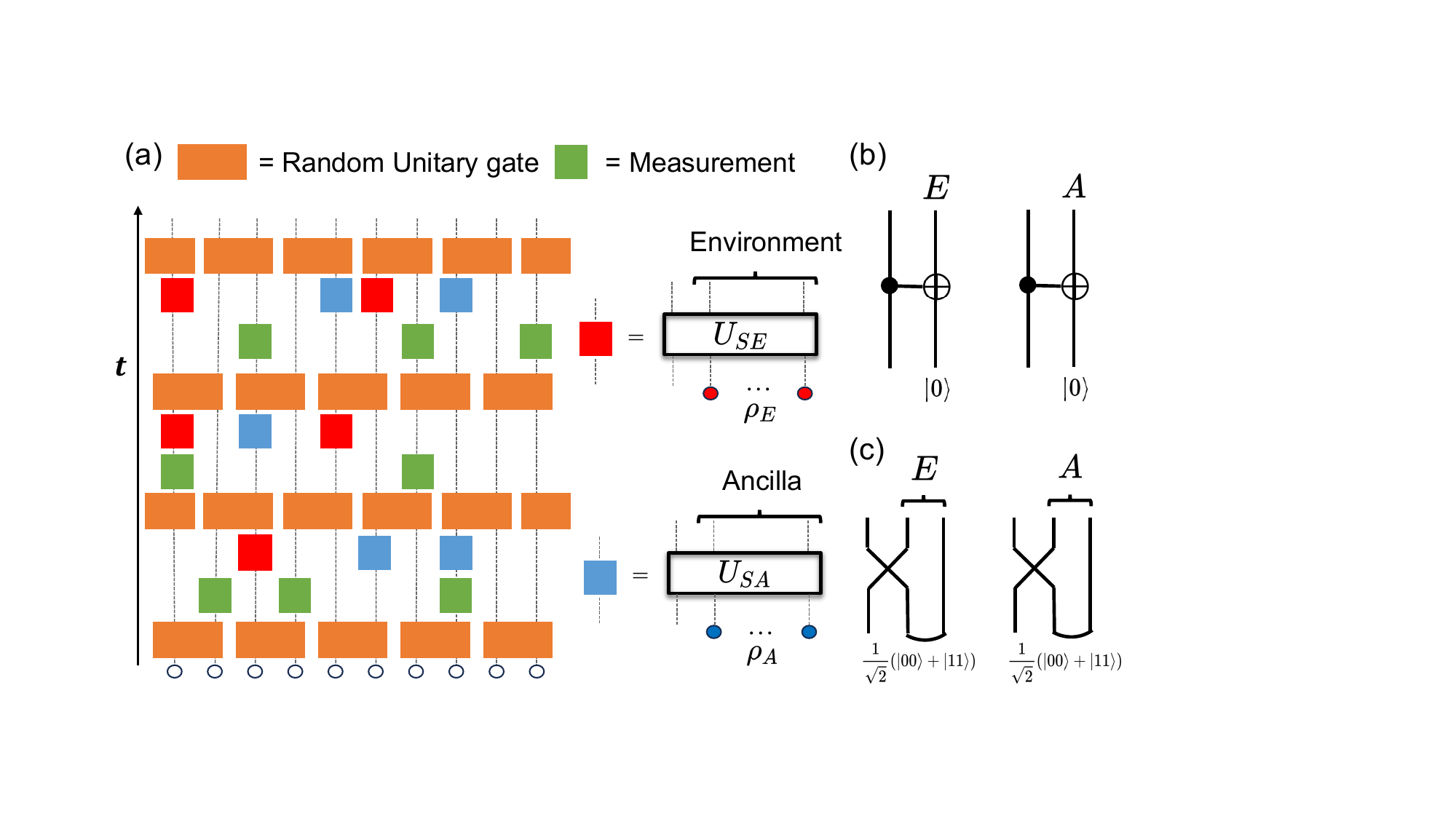}
\end{center}
\caption{Circuit structure and operations. (a) The quantum circuit model we consider in this work. Orange, green, red, and blue rectangles represent random unitary gates, measurements, noise, and QE operations, respectively. Noise and QE operation are further detailed as interactions with environment and ancilla qubits. (b) Dephasing noise and the corresponding symmetric QE operation. $U_{SE}=U_{SA}=\text{CNOT}$ and $\rho_E=\rho_A=\left | 0 \right \rangle \left \langle 0 \right |$. (c) Depolarizing noise and the corresponding symmetric QE operation. Notice that it requires two environment qubits to represent the noise. $U_{SE}=U_{SA}=\text{SWAP}$. $\rho_E$ and $\rho_A$ are in two-qubit Bell state.}
\label{fig1}
\end{figure}

\emph{Circuit model.}
We consider a quantum circuit where four types of operations are allowed: random unitary gate, measurement, noise, and QE operation. As shown in Fig.~\ref{fig1}(a), the unitary gates are applied in a brick wall pattern. Between every two unitary layers, each qubit has a probability $p$ of being measured projectively. Meanwhile, noise or QE operations occur with probabilities $q_{n}$ and $q_{e}$, respectively. Although noises are usually represented as quantum channels, we can adopt the Stinespring representation of the channel to represent it by unitary operations. Specifically, a quantum channel acting on the system can be written as  $\mathcal{N}(\rho) = \text{Tr}_{E}(U_{SE}(\rho \otimes\rho_E)U_{SE}^{\dagger}) $ where $U_{SE}$ represents the interaction between the system and environment, and $ \rho_{E} $ is the initial state of the environment~\cite{nielsen2010}. In the left column of Fig.~\ref{fig1}(b) and~\ref{fig1}(c), we show two examples of representing dephasing and depolarizing noise from this perspective. Similarly, QE operations can be considered as applying a unitary $U_{SA}$ to entangle the system with the ancilla qubits, whose initial states are $ \rho_{A} $. We provide representative examples in the right column of Fig.~\ref{fig1}(b) and~\ref{fig1}(c). It's worth noticing that the number of environment and ancilla qubits are not limited. Despite the similarity in the diagrams, it is crucial to emphasize that the environment qubits are always discarded at the end, signifying that the information is inevitably lost by tracing out these qubits. In contrast, the ancilla qubits remain under our control, allowing us to have full access to and manipulation of them. 

The input global state can be written as $\left | \Psi_{0}^{ESA} \right \rangle  = \left | \psi_{E} \right \rangle \otimes\left | \psi_{S} \right \rangle \otimes \left | \psi_{A} \right \rangle$ and a certain realization can be regarded as acting unitary gates and projection operators on it. At the end of every realization ${m}$, the global state can be represented as an unnormalized pure state $\left | \Psi_{m}^{ESA} \right \rangle$ and the physically relevant state is $\rho_{m}^{SA} = \text{Tr}_{E}\left(\left | \Psi_{m}^{ESA} \right \rangle\left \langle \Psi_{m}^{ESA}\right |\right) $. For convenience, we drop the superscript from now on. In this work, the realization $m$ includes both the circuit structure resulting from the random gates and operation locations, as well as the trajectory labeling by different measurement outcomes.

\begin{figure}[b]
\begin{center}
\includegraphics[width=3.4in, clip=true]{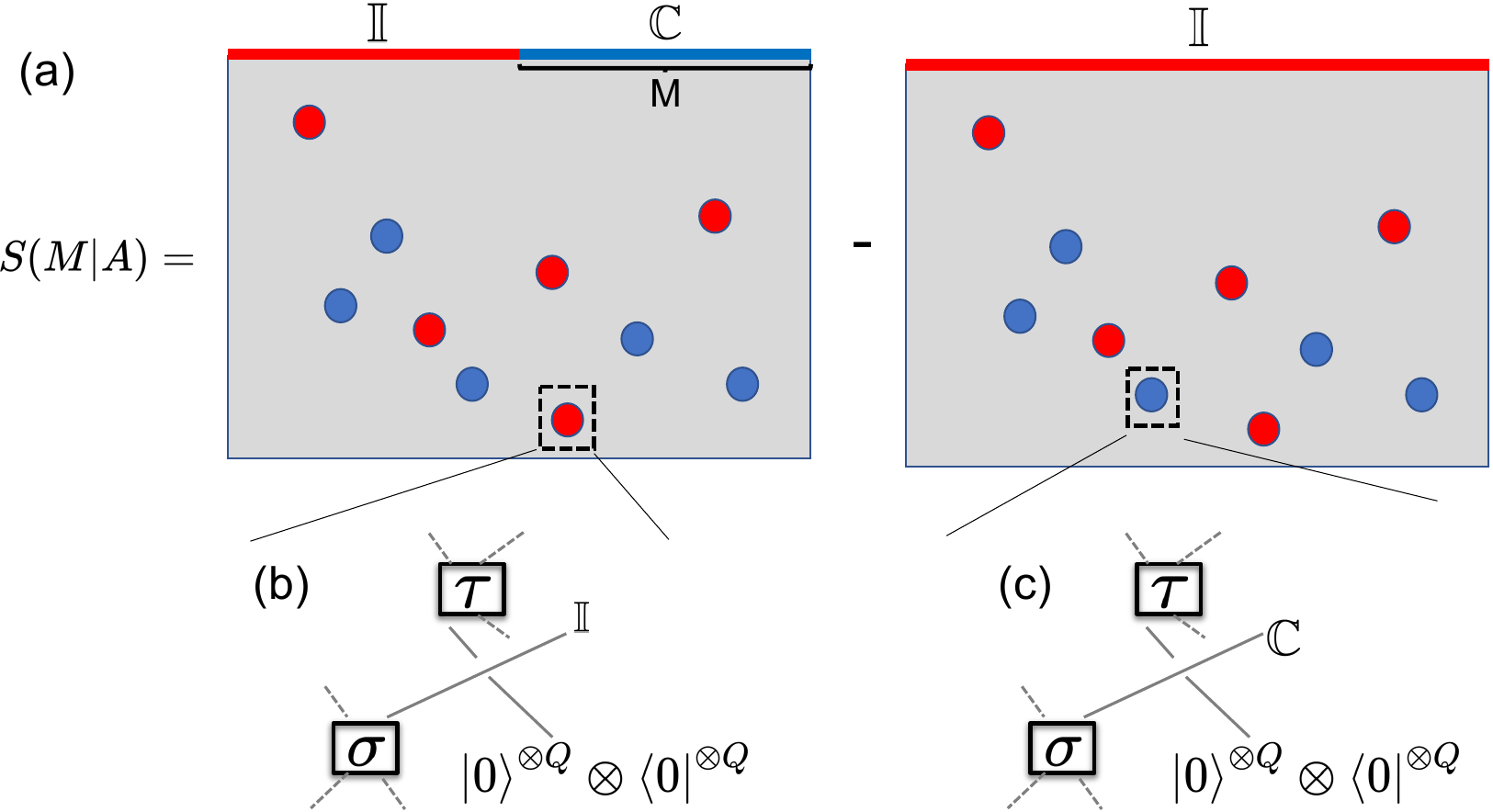}
\end{center}
\caption{Statistical-mechanics model. (a) $\overline{S(M|A)}$ is equal to the free energy difference of the same random field model under different boundary conditions. Red and blue color represents $\mathbb{I}$ and $\mathbb{C}$, respectively. The circles within the bulk indicate that the boundary conditions on environment and ancilla qubits are equivalent to applying extrinsic fields in different directions on the spins. Their locations are random due to the stochastic application of noise and QE operations. (b), (c) Detailed schematics illustrating the bond weight in the statistical-mechanics model, showing a qubit subjected to resetting noise and QE operations.}
\label{fig2}
\end{figure}

\emph{Analytical analysis.}
Instead of considering subsystem entanglement entropy, we aim to calculate the CEE $\overline{S(M|A)} \equiv \overline{S(M, A)} - \overline{S(A)}$. $M$ denotes an arbitrary subsystem and the overline here represents averaging all circuits and measurement outcomes. Mathematically, this can be expressed as,
\begin{eqnarray}
\overline{S(M|A)} &=& \lim_{n \to 1}(\overline{S^{(n)}(M,A)} -  \overline{S^{(n)}(A)}) \nonumber\\
&=&\lim_{n \to 1}\lim_{k \to 0}\frac{1}{(1-n)k}\text{log}\left(\frac{\mathcal{Z}^{(n,k)}_{MA}}{\mathcal{Z}^{(n,k)}_{A}}\right) \nonumber\\
&=&\lim_{n \to 1}\lim_{k \to 0}\frac{1}{(n-1)k}\left(\mathcal{F}^{(n,k)}_{MA}-\mathcal{F}^{(n,k)}_{A}\right),
\label{eq:one}
\end{eqnarray}
where
\begin{eqnarray}
\mathcal{Z}_{MA}^{(n,k)} &=& \sum_{m}\text{Tr}\left(\mathbb{C}^{M}\Lambda_{A}^{(n,k)}\right), \mathcal{Z}_{A}^{(n,k)} = \sum_{m}\text{Tr}\left(\Lambda_{A}^{(n,k)}\right),
\nonumber
\\
\Lambda_{A}^{(n,k)} &=& \text{Tr}_{A\cup E}\left[\left(\left | \Psi_m \right \rangle \left \langle \Psi_m \right |\right)^{\otimes Q}\mathbb{C}^{A}\right].
\label{eq:two}
\end{eqnarray}
$\mathcal{F}^{n,k}=-\text{log}\mathcal{Z}^{(n,k)}$ is identified as the free energy~\cite{supp}. The underlying statistical-mechanics model for $\mathcal{Z}^{(n,k)}$ consists of spins that take values in the permutation group $\mathbb{S}(Q)$ with $Q=nk+1$, where $\mathbb{I}$ and $\mathbb{C}$ correspond to two particular group element in $\mathbb{S}(Q)$~\cite{supp}. A critical observation is that, in both $\mathcal{Z}_{MA}$ and $\mathcal{Z}_{A}$, the environment qubits are subject to the same boundary condition $\mathbb{I}$, while the ancilla qubits are subject to $\mathbb{C}$. Consequently, we can interpret them as competing external fields applied at random positions within the bulk, with $\Lambda_{A}^{(n,k)}$ representing the bulk partition function. In other words, through QE operation and considering the conditional entropy, a symmetry-breaking field along the $\mathbb{C}$ direction is effectively introduced to counteract the influence of noise. Moreover, the unitary gates induce a ferromagnetic coupling, while the measurement rate acting as the temperature~\cite{supp}. Thus, $\overline{S(M|A)}$ is directly related to the free energy difference of a random field model with ferromagnetic coupling, subject to different boundary conditions, as shown in Fig.~\ref{fig2}(a). 

In the absence of random fields, the model exhibits spin permutation symmetry~\cite{symmetry}. However, any non-zero net external field explicitly breaks this symmetry. As a result, the bulk spins become aligned with either $\mathbb{I}$ or $\mathbb{C}$, leading to the formation of domain walls due to the imposed boundary conditions. The free energy cost associated with these domain walls scales proportionally with the length of the subsystem, resulting in a volume-law scaling of the CEE. A transition from volume-law to area-law scaling becomes possible only when the net external field vanishes, restoring the spin permutation symmetry between $\mathbb{I}$ and $\mathbb{C}$. 
This requires that the QE operation is symmetric to the noise, along with equal rates for them. Specifically, unless otherwise indicated, we assume the symmetric scenario where $U_{SE}=U_{SA}$, $\rho_E=\rho_A$ and $q_n = q_e = q/2$ in the following. Given prior work demonstrating phase transitions in random field Ising models and Potts models with zero-field conditon~\cite{ding2024a, binder1983, ding2024b, fontanari1989, kumar2022, nattermann1997, nishimori1983, rieger1993, aharony1976,fytas2013}, we anticipate that a phase transition would occur at finite $p$ and $q$ in quantum circuit. At high temperatures (large $p$), the system enters a paramagnetic phase where domain walls are absent, resulting in area-law scaling. Conversely, at low temperatures (small $p$), the symmetry is spontaneously broken and domain walls form along the boundary, yielding volume-law scaling.

\begin{figure}[t]
\begin{center}
\includegraphics[width=3.4in, clip=true]{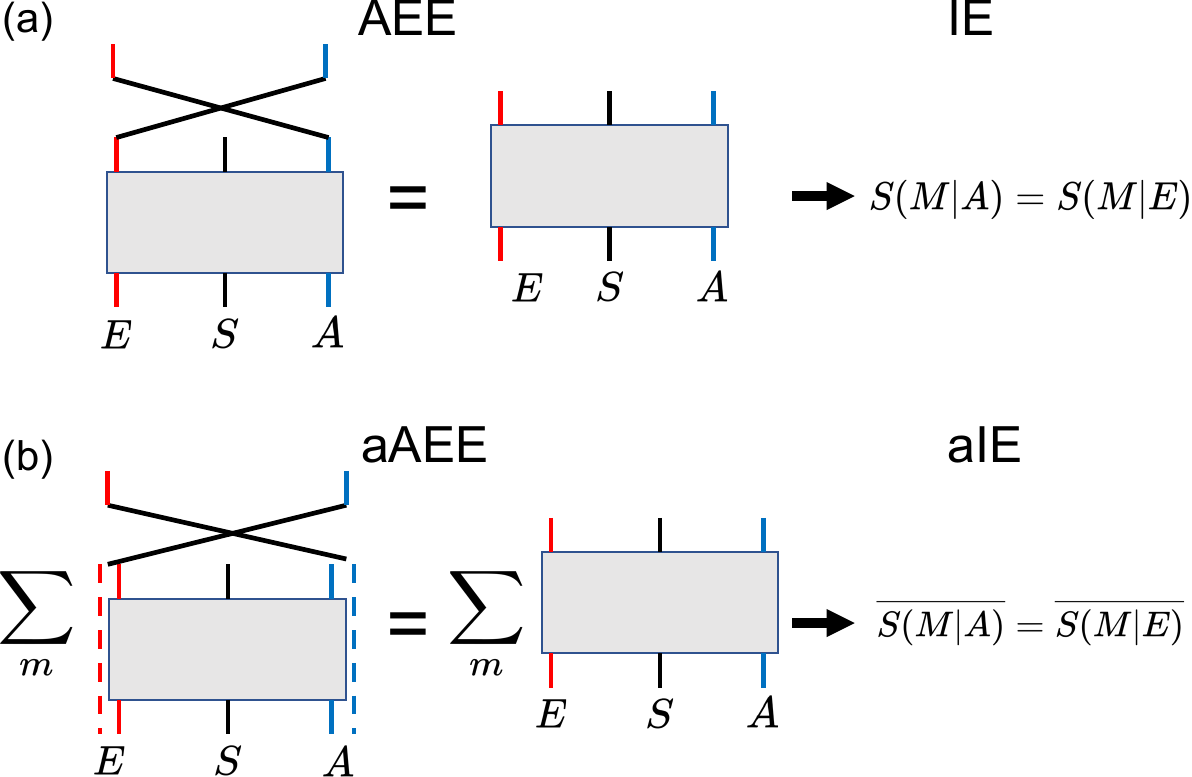}
\end{center}
\caption{Relationship between symmetries. (a) AEE symmetry requires that every realization remains invariant under the SWAP operation on environment and ancilla qubits, which consequently leads to IE symmetry. (b) The aAEE symmetry is the averaged version of AEE symmetry. The dashed lines indicate that unentangled qubits can be added to balance the number of qubits before applying the SWAP operation. The aAEE symmetry would then result in an aIE symmetry, which is the averaged version of IE symmetry.}
\label{fig3}
\end{figure}

\emph{Average symmetry.} Since $\overline{S(M|A)}$, as opposed to $\overline{S(M)}$ in traditional setting, undergoes a phase transition, it is pertinent to determine under what conditions $\overline{S(M|A)}$ serves as a valid probe of entanglement or quantum information~\cite{horodecki2009}. This ensures that the observed phase transition constitutes a genuine entanglement phase transition with meaningful operational significance. It has been proven that if the system has information exchange (IE) symmetry in every realization, such that $S(M|A)=S(M|E)$, then $S(M|A)$ ensures entanglement between $M$ and $M^{c}\cup A$~\cite{kelly2024b}. Here, $M^{c}$ denotes the complement of the subsystem $M$. Volume law scaling of $S(M|A)$ would then yield an extensive number of distillable Bell pairs between $M$ and $M^{c}\cup A$. The IE symmetry can be guaranteed if an apparatus-environment exchange (AEE) symmetry is present, which requires that each circuit realization be symmetric under the exchange of ancilla and environment qubits. However, this symmetry is explicitly violated by any realization in our circuit since noise and QE operation are independently chosen and applied at each location. Additionally, the number of ancilla and environment qubits may differ within a specific circuit realization, making the ``exchange'' operation more subtle to define. To solve this issue, we propose that an aAEE symmetry can still lead to an average IE (aIE) symmetry, thereby making the averaged CEE a valid probe of entanglement. The relationship between these symmetries are summarized in Fig.~\ref{fig3}. We first define a generalized SWAP operation $\text{SWAP}'$ on $\left | \Psi_m \right \rangle$ as initially adding unentangled qubits to either $E$ or $A$ to equalize the number of qubits. Subsequently, a conventional SWAP gate is applied to $E$ and $A$, followed by the removal of the unentangled qubits. $\text{SWAP}'\left | \Psi_m\right \rangle$ would not necessarily equal to $\left | \Psi_m \right \rangle$, but may transform it into another realization $\left | \Psi_{m'} \right \rangle$. The aAEE symmetry requires that the probabilities of these two realizations occurring are identical.  Concretely, one can represent the circuit’s outcome as $ \rho = \sum_{m}\left |\Psi_m \right \rangle \left \langle \Psi_m \right | $~\cite{aAEE}, and the aAEE symmetry necessitates that
\begin{equation}
\label{eq:three}
    \text{SWAP}'\rho\text{SWAP}'^{\dagger} = \rho.
\end{equation}
This is reminiscent of the recently proposed average (weak) symmetry in the study of mixed state order~\cite{fan2024, ma2023, ma2023c, sala2024, bao2023, ma2024, hauser2024}. With aAEE symmetry and combining with Eq.~(\ref{eq:two}), one would have 
\begin{eqnarray}
\sum_{m}\Lambda_A^{Q} &=& \text{Tr}_{A\cup E}(\rho^{\otimes Q}\mathbb{C}^{A})=\text{Tr}_{A\cup E}\left(\text{SWAP}'\rho^{\otimes Q}\text{SWAP}'^{\dagger}\right.
\nonumber\\
&&\left.\mathbb{C}^{A}\right)=\text{Tr}_{A\cup E}(\rho^{\otimes Q}\mathbb{C}^{E})\equiv \sum_{m}\Lambda^{Q}_{E}.
\end{eqnarray}
This result implies that $\overline{S(M|A)} = \overline{S(M|E)}$, which we call the aIE symmetry. It 
also ensures that $\overline{S(M|A)}$ represents the entanglement between $M$ and $M^{c}\cup A$.

Translating the aAEE symmetry into the concrete circuit model reveals that it actually coincides with the zero-field condition, which is necessary for satisfying Eq.~(\ref{eq:three}). Therefore, the MIPT captured by CEE indeed represents the transition in the entanglement scaling law of the underlying quantum state. Moreover, since any noise can be represented in the Stinespring form~\cite{nielsen2010}, QE operations are capable of protecting MIPT from various types of noise. 

\begin{figure}[t]
\begin{center}
\includegraphics[width=3.4in, clip=true]{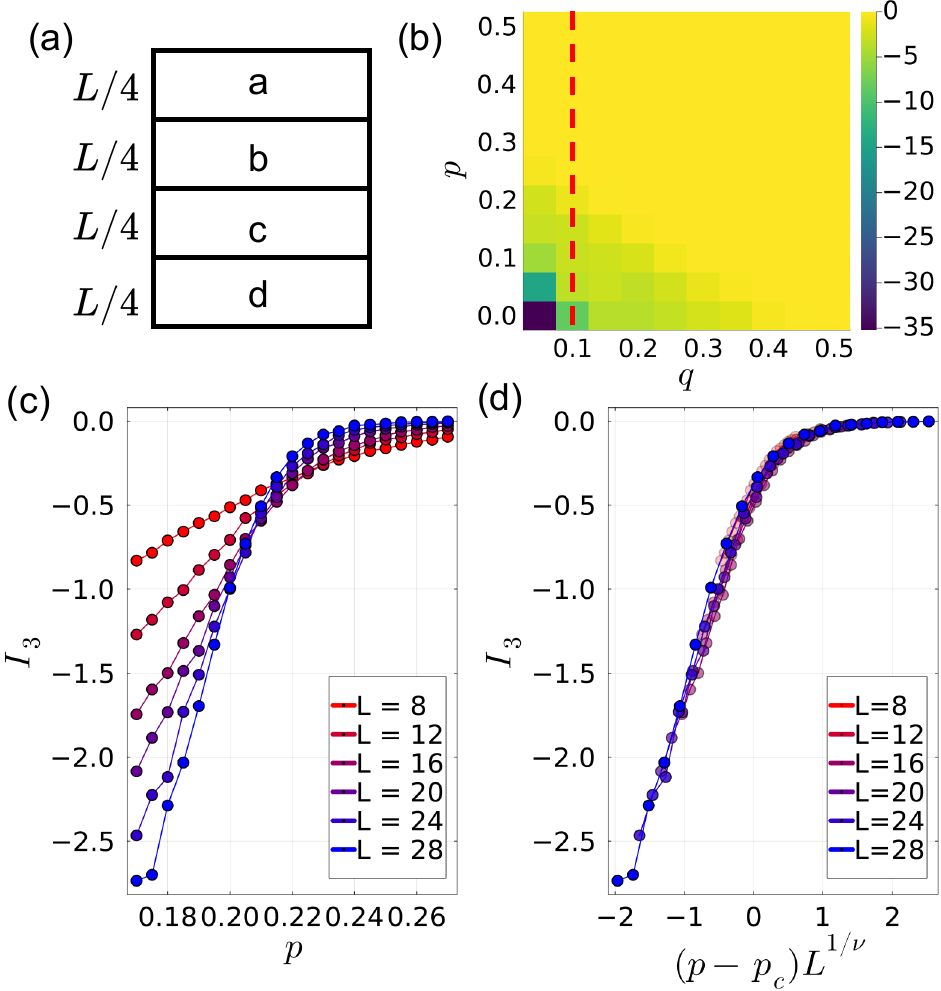}
\end{center}
\caption{Numerical results for dephasing noise. (a) We partition the system into four neighboring regions with equal size as a, b, c, d. The conditional tripartite mutual information is then calculated as $I_3 = S'(\text{a})+S'(\text{b})+S'(\text{c})-S'(\text{ab})-S'(\text{ac})-S'(\text{bc})+S'(\text{abc})$ where $S'(x)\equiv\overline{S(x|A)}$. (b) $I_3$ vs $(p,q)$ with $L=20$. The red line $q=0.1$ is further examined in (c), (d), which demonstrate MIPT in the presence of noise and QE operation. Each data point is averaged over $5\times10^4$ realizations.}
\label{fig4}
\end{figure}

\emph{Numerical results.}
We now turn to numerical calculations to explicitly demonstrate MIPT in the presence of noise and QE operations. Considering the absence of phase transitions in the 2D random field Ising model due to Imry-Ma argument~\cite{aharony1976}, we directly simulate the (2+1)-d quantum circuit with $L \times L$ qubits on a square lattice, corresponding to a 3D statistical model. To ensure efficient simulation, we choose the random unitary gates to be Clifford gates, allowing the circuit to be simulated using the stabilizer formalism~\cite{aaronson2004, gottesman1998, gottesman1997}. A single time step consists of four random unitary layers~\cite{supp}. Between every two layers, measurement in $z$-basis, noise, and QE operations occur at each site with probabilities $p$, $q/2$ and $q/2$, respectively. In the absence of any noise, MIPT occurs at around $p\sim0.3$. We evolve the circuit for depth $T=10L$ to achieve convergence in all simulations and employ the technique proposed in Ref.~\cite{kelly2024a} to perform efficient simulations without storing the ancilla qubits explicitly. In particular, we consider dephasing noise here, and results for other types of noise and additional numerical results can be found in Supplemental Material~\cite{supp}.

To determine the critical point, we consider the conditional tripartite mutual information $I_3$. We take periodic boundary condition in both directions and partition the system into four segments with equal size, as depicted in Fig.~\ref{fig4}(a). $I_3$ is expected to scale with $L$ in the volume law phase, while approach $0$ in the area law phase~\cite{lunt2021, sierant2022}. The complete phase diagram can thus be constructed by identifying the regions where $I_3$ is zero or not, as shown in Fig.~\ref{fig4}(b), where a volume law phase is clearly observed for non-zero values of $q$ and $p$. We further focus on the line $q=0.1$ and collapse the data according to the scaling form $I_3 = f((p-p_c)L^{1/\nu})$. The results are shown in Fig.~\ref{fig4}(c) and~\ref{fig4}(d). The critical point is $p_c=0.214(2)$ with the critical exponent $\nu = 0.9(1)$, which is close to the result for noiseless case in~\cite{sierant2022}. Details for data collapse  are provided in~\cite{supp}.

\emph{Discussions.} Similar to QEC, where more physical qubits are required to encode a smaller number of logical qubits, our scheme necessitates the preparation of additional ancilla qubits as the circuit depth increases. Specifically, $O(qNT)$ ancilla qubits are needed, where $N$ is the number of system qubits, $p$ is the noise rate, and $T$ is the circuit depth. Assuming $q_e\sim1\%, L=20$ and $T=10L$, around $800$ ancilla qubits are needed to observe the phase transition. It is imperative to underscore that we ignore potential noise in the ancilla qubits. This is justified, as each ancilla qubit is utilized only once and can be effectively isolated after use, thereby mitigating the risk of noise propagation and contamination.

An immediate application of our scheme requires prior knowledge of the probability that a specific type of decoherence noise will occur in the experimental setting. Our scheme can be directly adapted to determine the noise rate as follows: apply a finite amount of measurement such that the state enters the area law phase, then sweep through different QE operation rates $q_e$. As long as the circuit is deep in the area law phase, the bipartite CEE should not change upon varying $p$, once the condition $q_n=q_e$ is met. Therefore, curves of CEE corresponding to different $p$ would intersect at the point $q_n=q_e$ in the thermodynamic limit. By performing finite-size scaling on the intersection point, we obtain an estimate of the noise rate $q_n$ in the system. A more detailed explanation and finite size numerical results are provided in~\cite{supp}.

Besides noise, post-selection is another significant challenge preventing MIPT from being easily observed. Here the post-selection problem persists when the measurement rate is non-zero because determining the CEE of a single trajectory requires post-selecting on the measurement outcomes, a difficulty rooted in Born's rule. Recent work has utilized various post-selection-free methods to probe the phase transition, such as using cross-entropy benchmarking~\cite{li2023, tikhanovskaya2024} or providing bounds on the entanglement entropy~\cite{garratt2023a, mcginley2024}. We believe these methods can also be extended to the conditional entropy in this work, and we leave this exploration for future work.

Finally, we highlight the differences between our circuit model and that in Ref.~\cite{kelly2024a, kelly2024b}. Apart from incorporating measurement in our circuit, the most notable difference is that noise and QE operations act randomly and independently, not necessarily at the same location. The only requirement is that their rates are equal, which serves as a more realistic and feasible setting since we usually cannot accurately predict where and when noise will occur. Additionally, previous work has demonstrated a noise-induced phase transition where the noise rate is inversely proportional to the system size~\cite{liu2024a}. In our case, the noise rate is constant, which again represents a more realistic scenario.

\begin{acknowledgments}
\emph{Acknowledgment.} We thank Shane P. Kelly for valuable discussions. This work is supported by the Natural Science Foundation of China through Grants No.~12350404 and No.~12174066, the Innovation Program for Quantum Science and Technology through Grant No.~2021ZD0302600, the National Key Research Program of China under Grant No.~2019YFA0308404, the Science and Technology Commission of Shanghai Municipality under Grants No.~23JC1400600, No.~24LZ1400100 and No.~2019SHZDZX01.
\end{acknowledgments}


\begin{thebibliography}{85}
\bibitem [{sym()}]{symmetry}%
  \BibitemOpen
  \href@noop {} {\bibinfo {title} {Rigorously, the symmetry is $(\mathbb{S}_{Q}\times \mathbb{S}_{Q}) \rtimes \mathbb{Z}_2$.}}\BibitemShut {Stop}%
\bibitem [{aAE()}]{aAEE}%
  \BibitemOpen
  \href@noop {} {\bibinfo {title} {{{A subtlety in defining $\rho$ in this manner is that the number of qubits in $\left | \Psi_m \right \rangle$ may vary, but this can be simply addressed by adding unentangled qubits to states with fewer qubits. }}}}\BibitemShut {NoStop}%
\end{thebibliography}
\end{document}


\title{Supplemental Material for ``Protect Measurement-induced Phase Transition from Noise''}
\author{Dongheng Qian}
\affiliation{State Key Laboratory of Surface Physics and Department of Physics, Fudan University, Shanghai 200433, China}
\affiliation{Shanghai Research Center for Quantum Sciences, Shanghai 201315, China}
\author{Jing Wang}
\thanks{wjingphys@fudan.edu.cn}
\affiliation{State Key Laboratory of Surface Physics and Department of Physics, Fudan University, Shanghai 200433, China}
\affiliation{Shanghai Research Center for Quantum Sciences, Shanghai 201315, China}
\affiliation{Institute for Nanoelectronic Devices and Quantum Computing, Fudan University, Shanghai 200433, China}
\affiliation{Hefei National Laboratory, Hefei 230088, China}


\maketitle

\tableofcontents

\section{Analytic mapping}
\begin{figure}[t]
\begin{center}
\includegraphics[width=6in, clip=true]{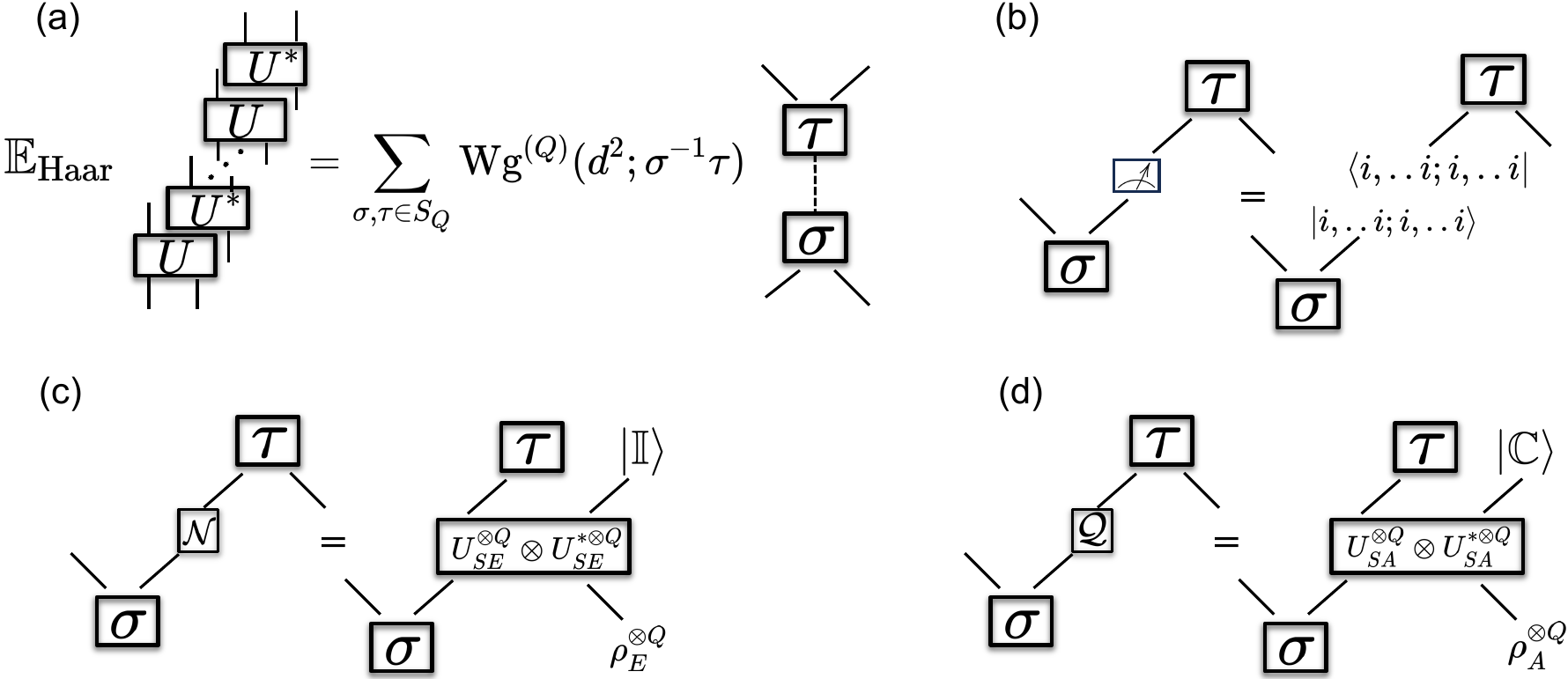}
\end{center}
\caption{Diagrammatic representations of operations in the circuit. (a) Random unitary gate. Each 2-qudit unitary gate, sampled from the Haar measure, can be simply represented as two spins, $\sigma$ and $\tau$ connected by a vertical bond with the bond weight given by the Weingarten function. (b) Projective measurement. For the measurement outcome $i$, each replica is acted upon by the same projector $\left | i \right \rangle \left \langle i \right |$. (c) Noise. The environment qubits are traced out, equivalent to taking the inner product with $\left | \mathbb{I} \right \rangle$. (d) QE operation. Since we consider the CEE, the ancilla qubits are equivalently taken inner product with $\left | \mathbb{C} \right \rangle$.}
\label{fig1}
\end{figure}
In this section, we provide a detailed explanation of how  the calculation of conditional entanglement entropy (CEE) can be mapped to the free energy difference of a statistical mechanics model under different boundary conditions. We further show the derivation of the bond weights under various noises and the corresponding symmetric QE operations. Our derivation is largely based on Refs.~\cite{bao2020, jian2020a, zhou2019, dias2023, liu2023e, weinstein2022}.

For clarity, we separately label the measurement trajectory  by $m$ and denote $\rho_m = \left | \Psi_m \right \rangle \left \langle \Psi_m \right |$ as the purified state with support on the environment, system and ancilla qubits corresponding to the trajectory $m$. This state is unnormalized, with the Born probability of occurrence given by $p_m = \text{Tr}(\rho_m)$. By definition, we have:
\begin{eqnarray}
 \overline{S(M|A)} = \lim_{n \to 1}\overline{S^{(n)}(M|A)} = \lim_{n \to 1}\mathop{\mathbb{E}}\limits_{\text{circuits}} \sum_{m}p_m\frac{1}{1-n}\text{log}\left[\frac{\text{Tr}(\rho_{MA,m}^{n})}{\text{Tr}(\rho_{A,m}^n)}\right]
\end{eqnarray}
where $ \rho_{MA, m} = \text{Tr}_{M^{c}\cup E}(\rho_m)$ and $\rho_{A,m}=\text{Tr}_{S\cup E}(\rho_m)$. We denote $\mathbb{S}$ to be the cyclic permutation operation in permutation group $\mathbb{S}(n)$, and define its action on the $n$-fold replicated single qubit Hilbert space as:
\begin{equation}
\mathbb{S} = \sum_{i_1, i_2, ..i_n}\left | i_2,i_3, ..i_n,i_1\right \rangle \left \langle i_1,i_2, ..i_{n-1},i_n\right |.
\end{equation}
We can further define $\mathbb{S}^{X} = \otimes_{i \in X}\mathbb{S}_i$. Using the identities $ \text{Tr}(\rho_{X,m}^{n}) = \text{Tr}(\rho_m^{\otimes n}\mathbb{S}^{X})$ and $ \text{log}x=\lim_{k \to 0} \frac{x^{k}-1}{k}$, we obtain:
\begin{eqnarray}
\overline{S(M|A)}=&&\lim_{n \to 1}\lim_{k \to 0}\mathop{\mathbb{E}}\limits_{\text{circuits}}\sum_{m}\frac{1}{(1-n)k}\text{Tr}(\rho_m)\left[\text{Tr}(\rho_m^{\otimes n}\mathbb{S}^{MA})^{k}-\text{Tr}(\rho_m^{\otimes n}\mathbb{S}^{A})^{k}
\right] \nonumber \\ 
=&& \lim_{n \to 1}\lim_{k \to 0}\mathop{\mathbb{E}}\limits_{\text{circuits}}\sum_{m}\frac{1}{(1-n)k}\left[\text{Tr}(\rho_m^{\otimes Q}\mathbb{C}^{MA}) - \text{Tr}(\rho_m^{\otimes Q}\mathbb{C}^{A})\right].
\end{eqnarray}
Here, $\mathbb{C}^{X} = \otimes_{i \in X}(\mathbb{S}^{\otimes k}\otimes I)$ and $Q=nk+1$. We can define $\Lambda_{A}^{Q}=\text{Tr}_{A\cup E}[\rho_m^{Q}\mathbb{C}^{A}]$, and the above equation eventually becomes:
\begin{eqnarray}
\overline{S(M|A)}=&&\lim_{n \to 1}\lim_{k \to 0}  \mathop{\mathbb{E}}\limits_{\text{circuits}}\sum_{m}\frac{1}{(1-n)k}[\text{Tr}_{S}(\Lambda_{A}^{Q}\mathbb{C}^{M}) - \text{Tr}_{S}(\Lambda_{A}^{Q})] = \lim_{n \to 1}\lim_{k \to 0} \frac{1}{(1-n)k}(\mathcal{Z}_{MA}^{(Q)}-\mathcal{Z}_{A}^{(Q)}) \\
&&\mathcal{Z}_{MA}^{(Q)} =\mathop{\mathbb{E}}\limits_{\text{circuits}}\sum_{m}\text{Tr}_{S}\left(\mathbb{C}^{M}\Lambda_{A}^{(Q)}\right), \mathcal{Z}_{A}^{(Q)} = \mathop{\mathbb{E}}\limits_{\text{circuits}}\sum_{m}\text{Tr}_{S}\left(\Lambda_{A}^{(Q)}\right).
\end{eqnarray}
Notice that $\lim_{k\to 0}\mathcal{Z}_{MA}^{(Q)} = \mathcal{Z}_{MA}^{(1)}=1$ and $\lim_{k\to 0}\mathcal{Z}_{A}^{(Q)} = \mathcal{Z}_{A}^{(1)}=1$. Therefore, we have:
\begin{equation}
\lim_{k \to 0}(\mathcal{Z}_{MA}^{(Q)}-\mathcal{Z}_{A}^{(Q)}) = \lim_{k\to0}\text{log}\left(\frac{\mathcal{Z}_{MA}^{(Q)}}{\mathcal{Z}_{A}^{(Q)}}\right)
\end{equation}
which recovers the results in the main text. We can further employ a convenient vector notation where the replicated density matrix $\Lambda_{A}^{Q}$ is regarded as a vector state $\left | \Lambda_{A}^{Q} \right \rangle$ in the $ \mathcal{H}^{\otimes Q}\otimes \mathcal{H}^{*\otimes Q}$ Hilbert space. We define states corresponding to a particular group member $g$ in the permutation group $S(Q)$ as:
\begin{equation}
\left | g \right \rangle = \sum_{i_1,i_2,...i_{Q}}^{d}\left | i_{g(1)},i_{g(2)}, ...i_{g(Q)};i_{1},i_{2},...i_{Q} \right \rangle
\label{eq:S8}
\end{equation}
where $d$ is the local Hilbert dimension. Thus, we have 
$\mathcal{Z}_{MA}^{Q} =  \left \langle \mathcal{Z}_{\text{bulk}} |\ \mathbb{C}^{M}\otimes \mathbb{I}^{M^{c}} \right \rangle$
and
$\mathcal{Z}_{A}^{Q} =  \left \langle \mathcal{Z}_{\text{bulk}} |\ \mathbb{I}^{M\cup M^{c}} \right \rangle$
\normalsize
with $\left | \mathcal{Z}_{\text{bulk}} \right \rangle = \mathop{\mathbb{E}}\limits_{\text{circuits}}\sum_m\left |\Lambda_{A}^{(Q)} \right \rangle$. The  goal is then to represent $\left | \mathcal{Z}_{\text{bulk}} \right \rangle$ as a statistical mechanics model for spins taking value in permutation group $\mathbb{S}_{Q}$, with $\mathcal{Z}_{MA}^{Q}$ and $\mathcal{Z}_{A}^{Q}$ being the partition function with the boundary conditions $\left | \mathbb{C}^{M}\otimes \mathbb{I}^{M^{c}}\right \rangle$ and $\left | \mathbb{I}^{M\cup M^{c}} \right \rangle$, respectively.

With the notation in Eq.~(\ref{eq:S8}), we can represent various operations in the circuit in an intuitive and simple manner. Two identities are particularly useful: $\left \langle \sigma | i,..i;i,..i\right \rangle = 1 $ and $\left \langle \sigma | \tau \right \rangle = d^{|\sigma \tau^{-1}|}$, where $|g|$ denotes the number of cycles in permutation $g$. It’s evident to see that this inner product reaches its maximum when $\sigma = \tau$, which suggests that it acts like $d^{Q}\delta_{\sigma, \tau}$ in the limit $d \to \infty$. The random unitary gates contribute to the vertical bond weight, given by:
\begin{equation}
\mathbb{E}_{\text{Haar}}[U_{ij}^{\otimes Q}\otimes U_{ij}^{*\otimes Q}] = \sum_{\{\sigma, \tau\}\in \mathbb{S}(Q)}\text{Wg}^{Q}(d^{2};\sigma^{-1}\tau)\left | \sigma\sigma\right \rangle_{ij}\left \langle \tau \tau \right |_{ij}
\end{equation}
as depicted in Fig.~\ref{fig1}(a). Here, $\text{Wg}^{Q}(d^{2};g)$ is the Weingarten function for the unitary group $U(d^{2})$~\cite{collins2006}. In the $d \to \infty$ limit, it behaves like:
\begin{equation}
\lim_{d \to \infty} \text{Wg}^{(Q)}(d^2;\sigma^{-1}\tau)\sim\frac{\text{Moeb}(\sigma^{-1}\tau)}{d^{4Q-2|\sigma^{-1}\tau|}}.
\end{equation}
where $\text{Moeb}(\sigma^{-1}\tau)$ is the Moebius number of permutation $\sigma^{-1}\tau$. Thus, it acts as the ferromagnetic coupling along the vertical bonds which aligns $\sigma$ and $\tau$.
Other operations correspond to the weight $W$ on non-vertical bonds. Consequently, the statistical mechanics model is an anisotropic $Q!$ state model on a honeycomb lattice, with the partition function given by:
\begin{equation}
\mathcal{Z}_{\text{bulk}} = \sum_{\{\sigma,\tau\} \in S(Q)}\prod_{\left \langle\sigma \tau\right \rangle \in {V^{c}}}W(\sigma, \tau)\prod_{\left \langle\sigma \tau\right \rangle \in {V}}\text{Wg}^{(Q)}(d^{2};\sigma^{-1}\tau)
\end{equation}
where $V$ denotes all the vertical bonds and $V^{c}$ denotes the non-vertical bonds.

For a projective measurement with outcome $i$, the weight is $\left \langle \sigma | i,..i;i,..i\right \rangle \left \langle i,..i;i,..i | \tau \right \rangle = 1$, as shown in Fig.~\ref{fig1}(b). Since we further average over all the measurement outcomes, the measurement contributes a weight $d$ on the bond. Noise and QE operations can also be depicted with corresponding pictorial representation as shown in Figs.~\ref{fig1}(c) and~\ref{fig1}(d). Although obtaining a simple analytical expression for the bond weight for certain types of noise and symmetric QE operation is generally difficult, simple expressions exist for both resetting noise and depolarizing noise. For resetting noise and the symmetric QE operation, where $U_{SE} = U_{SA} = \text{SWAP}$ and $\rho_E = \rho_A = \left | 0 \right \rangle \left \langle 0 \right |$, one can easily identify from the diagram that the weight for noise is $\left \langle \sigma | \mathbb{I} \right \rangle \left \langle 0,..0;0,..0|\tau \right \rangle = \left \langle \sigma | \mathbb{I} \right \rangle $, and the weight for symmetric QE operation is similarly $\left \langle \sigma | \mathbb{C} \right \rangle $. Thus, the non-vertical bond weight is:
\begin{eqnarray}
    W^{\text{reset}}(\sigma, \tau) =&&(1-p)(1-q)\left \langle \sigma | \tau \right \rangle 
     + (1-p)\frac{q}{2}(\left \langle \sigma | \mathbb{C} \right \rangle +\left \langle \sigma | \mathbb{I} \right \rangle) + pd.
\label{eq:S11}
\end{eqnarray}
For depolarizing noise, with $U_{SE}=U_{SA}=\text{SWAP}, \rho_{E}=\rho_{A}=\frac{1}{\sqrt{2}}(\left | 00 \right \rangle + \left | 11 \right \rangle)(\left \langle 00 \right | + \left \langle 11 \right |) $, the result can be similarly derived as:
\begin{eqnarray}
    W^{\text{depo}}(\sigma, \tau) =(1-p)(1-q)\left \langle \sigma | \tau \right \rangle + (1-p)\frac{q}{2}(\left \langle \sigma | \mathbb{C} \right \rangle\left \langle \mathbb{C} | \tau  \right \rangle  +\left \langle \sigma | \mathbb{I} \right \rangle\left \langle \mathbb{I} | \tau  \right \rangle)
    + dp\frac{q}{2}(\left \langle \sigma | \mathbb{I}  \right \rangle + \left \langle \sigma | \mathbb{C}  \right \rangle) + p(1-q)d. 
\label{eq:S12}
\end{eqnarray}
Thus, it is evident that noise and symmetric QE operation act as  two competing external fields applied at random locations in both cases. For dephasing noise and symmetric QE operation, where $U_{SE}=U_{SA}=\text{CNOT}$ and $\rho_E=\rho_A=\left | 0 \right \rangle \left \langle 0 \right |$, the situation is more complicated. We have:
\begin{eqnarray}
&&\left \langle \sigma | \mathcal{N} | \tau \right \rangle = \sum_{\{i,\bar{i},j, \bar{j}\}}\prod_{l=1}^{Q}\delta_{i_{l},\bar{i}_{\sigma(l)}}\prod_{l=1}^{Q}\delta_{i_l,j_l}\delta_{\bar{i}_l,\bar{j}_l}\prod_{l=1}^{Q}\delta_{j_l,\bar{j}_l}\prod_{l=1}^{Q}\delta_{i_l,\bar{i}_{\tau(l)}} = \sum_{\{i\}}\prod_{l=1}^{Q}\delta_{i_l,i_{\sigma(l)}}\prod_{l=1}^{Q}\delta_{i_l,i_{\tau(l)}}, \\
&&\left \langle \sigma | \mathcal{Q} | \tau \right \rangle = \sum_{\{i,\bar{i},j, \bar{j}\}}\prod_{l=1}^{Q}\delta_{i_{l},\bar{i}_{\sigma(l)}}\prod_{l=1}^{Q}\delta_{i_l,j_l}\delta_{\bar{i}_l,\bar{j}_l}\prod_{l=1}^{Q}\delta_{j_l,\bar{j}_{\mathbb{C}(l)}}\prod_{l=1}^{Q}\delta_{i_l,\bar{i}_{\tau(l)}} = \sum_{\{i\}}\prod_{l=1}^{Q}\delta_{i_l,i_{\sigma\mathbb{C}^{-1}(l)}}\prod_{l=1}^{Q}\delta_{i_l,i_{\tau\mathbb{C}^{-1}}(l)},
\end{eqnarray}
where ${i_l, j_l}$ denotes the system and environment/ancilla qubit in the $l$-th replica. The terms $\prod_{l=1}^{Q}\delta_{i_l,j_l}\delta_{\bar{i}_l,\bar{j}_l}$ arise from the $\text{CNOT}$ gate, and $\prod_{l=1}^{Q}\delta_{j_l,\bar{j}_l}, \prod_{l=1}^{Q}\delta_{j_l,\bar{j}_{\mathbb{C}(l)}}$ represent inner product with $\mathbb{I}$ and $\mathbb{C}$, respectively.  Although a further simplified expression is unknown, it’s clear that
\begin{eqnarray}
\left \langle \sigma | \mathcal{N} | \mathbb{I} \right \rangle =&& \left \langle \mathbb{I} | \mathcal{N} | \sigma \right \rangle = \left \langle \sigma | \mathcal{N} | \sigma \right \rangle = \left \langle \sigma |  \mathbb{I} \right \rangle = d^{|\sigma|} \\
\left \langle \sigma | \mathcal{Q} | \mathbb{C} \right \rangle =&& \left \langle \mathbb{C} | \mathcal{Q} | \sigma \right \rangle = \left \langle \sigma | \mathcal{Q} | \sigma \right \rangle = \left \langle \sigma |  \mathbb{C} \right \rangle = d^{|\sigma\mathbb{C}^{-1}|}.
\end{eqnarray}
Meanwhile, other inner products $\left \langle \sigma | \mathcal{N} | \tau \right \rangle$ and $\left \langle \sigma | \mathcal{Q} | \tau \right \rangle$ are all smaller by at least a factor of $1/d$. Therefore, we have:
\begin{eqnarray}
    W^{\text{deph}}(\sigma, \tau) \simeq (1-p)(1-q)\left \langle \sigma | \tau \right \rangle 
     + (1-p)\frac{q}{2}(\left \langle \sigma | \mathbb{C} \right \rangle +\left \langle \sigma | \mathbb{I} \right \rangle)\delta_{\sigma,\tau} + pd 
\label{eq:S17}
\end{eqnarray}
in the $d \to \infty$ limit, where the two competing random fields $\mathbb{C}$ and $\mathbb{I}$ are again present. We conjecture that for $U=U_{SE}=U_{SA}$ acting non-trivially on both system qubit and the environment/ancilla qubit, noise and symmetric QE operation will always map to the extrinsic fields in the directions of $\mathbb{I}$ and $\mathbb{C}$ in the limit $d \to \infty$.

We now further examine the symmetry of the random field model with non-vertical bond weight in Eq.~(\ref{eq:S11}), Eq.~(\ref{eq:S12}) and Eq.~(\ref{eq:S17}). The symmetry group of the statistical mechanics model in the absence of noise and QE operation is $ (\mathbb{S}_{Q}\times \mathbb{S}_{Q}) \rtimes \mathbb{Z}_2$. The $\mathbb{S}_{Q}\times \mathbb{S}_{Q}$ symmetry is from the invariance of the weights under transformation $ \sigma \to g_i\sigma g_j$, where $g_i, g_j \in \mathbb{S}_{Q}$. The $\mathbb{Z}_2$ symmetry $\sigma \to \sigma^{-1}$ is due to the hermiticity of the density matrix.  When random fields are present, the original $\mathbb{S}_{Q}\times \mathbb{S}_{Q}$ symmetry is broken and the symmetry group becomes $ \mathcal{C}_{\mathbb{S}_{Q}}(\mathbb{C})\rtimes \mathbb{Z}_{2}$. The $\mathbb{Z}_2$ symmetry remains to be $\sigma \to \sigma^{-1}$. The group $\mathcal{C}_{\mathbb{S}_{Q}}(\mathbb{C})$ is the subgroup of $\mathbb{S}_{Q}$ that stabilizes $\mathbb{C}$, meaning that any group element $g$ satisfies $[g, \mathbb{C}] = 0$. The corresponding symmetry operation is then $\sigma \to g\sigma g^{-1}$. Furthermore, if we impose the condition of zero net random field such that $q_e=q_n=q/2$,  an additional $\mathbb{Z}_2$ symmetry arises, corresponding to the operation $\sigma \to \sigma^{-1} \mathbb{C}$. The full symmetry group is $ (\mathcal{C}_{\mathbb{S}_{Q}}(\mathbb{C})\rtimes \mathbb{Z}_2)\rtimes \mathbb{Z}_{2}$. MIPT under noise and QE operation can thus be understood as the $ (\mathcal{C}_{\mathbb{S}_{Q}}(\mathbb{C})\rtimes \mathbb{Z}_2)\rtimes \mathbb{Z}_{2}$ symmetry spontaneously breaks into the $ \mathcal{C}_{\mathbb{S}_{Q}}(\mathbb{C})\rtimes \mathbb{Z}_{2}$ symmetry.

To summarize, we have shown that under various noises, the calculation of conditional entanglement entropy can be mapped to the free energy difference of a  random field model under different boundary conditions. Although we are ultimately interested in the limit $Q\to 1$ and finite local Hilbert space dimension $d$, certain limiting cases can provide useful insights. One such case is the limit $d \to \infty$, where the model transforms into a random field $Q!$-state Potts model. Evidence suggests that when $Q$ is larger than $2$, the lower critical dimension is $D_c=2$~\cite{fontanari1989, kumar2022, nishimori1983, aharony1976,kumar2018}. While it is well established that the $Q!$-state Potts model without random fields becomes a bond percolation problem in the limit $Q\to 1$~\cite{fortuin1972}, the behavior in the presence of competing random fields remains unknown. Another useful limiting case is $Q=2$, where the model becomes a random field Ising model even for finite $d$. The conclusion is similar that the lower critical dimension is $D_c = 2$~\cite{rieger1993, nattermann1997, binder1983, fytas2013}. Based on these observations, we anticipate that the lower critical dimension for $Q \to 1$ and finite $d$ is also $2$, necessitating a $(2+1)$-d quantum circuit. Interestingly, we have also conducted numerical simulation for the $(1+1)$-d case and found evidence of a phase transition at a much lower rate of noise. These findings suggest that in the actual $d=2$ setting, the combined effects of noise and quantum-enchanced (QE) operations exhibit a less pronounced impact compared to the influence of two independent competing random fields.

\section{Entanglement probe}
In this section, we further elaborate on how the average information exchange (aIE) symmetry qualifies CEE as a valid probe of entanglement in the presence of noise and QE operations. 

Entanglement is a property of a composite system that cannot be separated into its smaller subsystems~\cite{horodecki2009}. This is directly related to the fact that, unlike classical system, knowledge of the entire entangled state does not necessarily provide complete information about its subsystems. A straightforward method to check for entanglement is to see if the state violates the inequality $S(A|B) = S(AB) - S(B) > 0$, where $S$ is the von Neumann entropy, which is non-negative by definition. A negative $S(A|B)$, which is equivalent to a non-zero $S(B)$ when the entire state is pure, would directly violate this inequality. For pure states, violation and inseparability are necessary and sufficient conditions for each other. However, the situation becomes more subtle for mixed state. The requirement of violation becomes too stringent, such that there are entangled states that do not violate the inequality. Nonetheless, a negative $S(A|B)$ remains a sufficient condition to ensure that $A$ and $B$ are entangled. This has further operational meaning, as a negative $S(A|B)$ implies the ability to distill $|S(A|B)|$ pairs of qubits in maximally entangled states~\cite{horodecki2005, horodecki2006}. 

We now show in detail how the aIE symmetry $\overline{S(M|A)} = \overline{S(M|E)}$ guarantees $\overline{S(M|A)}$ as a valid entanglement probe in detecting MIPT. The proof is similar to that for IE symmetry~\cite{kelly2024b}. It follows straightforwardly that $\overline{S(MA)} = \overline{S(M^{c}E)} = \overline{S(M^{c}A)}$, utilizing the purity of the total state. This leads to $\overline{S(M|A)} = \overline{S(M^{c}|A)}$. Additionally, we have:
\begin{equation}
\overline{S(M|A\cup M^{c})} = \overline{S(E)} - \overline{S(EM)} = -\overline{S(M|E)} = -\overline{S(M|A)}
\end{equation}
where the first equation uses the purity of the state, and the last equation uses the aIE symmetry. Finally, applying the sub-additivity of the von Neumann entropy, we obtain:
\begin{equation}
\overline{S(A)} + \overline{S(E)} \leq \overline{S(M^{c}E)} + \overline{S(ME)} = \overline{S(MA)}+ \overline{S(ME)} \Leftrightarrow \overline{S(M|A)} \geq 0.
\end{equation}
Thus, $\overline{S(M|A)}$ almost plays the same role as the subsystem entanglement entropy in pure states. Any non-zero $\overline{S(M|A)}$ implies a negative $\overline{S(M|M^{c}\cup A)}$, ensuring the entanglement between $M$ and $M^{c}\cup A$. The only caveat is that a zero $\overline{S(M|A)}$ does not necessarily rule out the possibility of entanglement between $M$ and $M^{c}\cup A$. This is similar to the entanglement negativity $E_{N}$, which is also non-negative, and $E_{N} =0$ does not guarantee that the state is separable. In the study of MIPT, we are interested in how the entanglement scales with the subsystem size, rather than whether there is a finite entanglement. Specifically, $\overline{S(M|A)}$ is always strictly greater than $0$ as long as $p\neq 1$, making this subtle point irrelevant when discussing the volume law to area law phase transition. Based on these reasons, the aIE symmetry indeed ensures that CEE is a valid probe of entanglement in the presence of noise.

\section{Numerical details}
\subsection{Efficient simulation}
In this section, we introduce the method we used in numerical calculation. We first review the stabilizer formalism used to achieve efficient numerical simulation~\cite{aaronson2004, gottesman1998, gottesman1997}. We denote the Pauli group on $n$ qubits, $G_n$, as the group consisting of all Pauli matrices ${I, X, Y, Z}$ along with multiplicative factors ${\pm1, \pm i}$. A stabilizer group $S$ is a subgroup of $G_n$ generated by elements $g_1,...,g_l$. These generators are independent commuting generators. We can associate a stabilizer state with every stabilizer group as: 
\begin{equation}
\rho = \sum_{g \in G_n}g = \frac{2^l}{2^{n}}\prod_{i=1}^{l}\frac{1+g_i}{2}
\end{equation}
When $n=l$, the stabilizer state becomes a pure state. These states can be represented efficiently by associating two $l \times n$ binary matrices, assuming phase is not considered. Specifically, each stabilizer can be represented as $g = \prod_{i=1}^{n}X_i^{\alpha_i}Z_i^{\beta_i}$ up to a phase factor, where $\alpha_i, \beta_i$ are binary numbers taking value in ${0, 1}$. Thus, instead of storing $2^n$ complex numbers, any stabilizer state can be efficiently represented by only $O(n)$ numbers. 

We demonstrate that all operations in the circuit map states to other states with a stabilizer representation, enabling efficient simulation. The two-qubit unitary gates are randomly sampled from the Clifford group, which is generated by $\{\text{CNOT}, \text{SWAP}, \text{H}, \text{P}\}$, representing the CNOT, SWAP, Hadamard and phase gate, respectively. According to the Gottesman–Knill theorem~\cite{nielsen2010}, a quantum circuit consisting solely of Clifford gates and measurements represented by Pauli operators can be efficiently simulated.  Specifically,  Clifford gates have the property of being the normalizer of $G_n$, meaning that $g' = UgU^{\dagger}$ still lies within the stabilizer group. Thus, we can track the evolution of the stabilizers rather than the state itself. For a measurement represented by a Pauli operator $g$, there are three possibilities:
\begin{itemize}
\item $g$ is in the stabilizer. The measurement result is determined by the phase, and the state is unchanged. 
\item $g$ is not in the stabilizer but commutes with all stabilizers. The measurement outcome is random, and $g$ is added to the stabilizer generators.
\item $g$ is not in the stabilizer and anti-commutes with a single generator $g'$. The measurement outcome is random, and the original $g'$ is replaced by $g$.
\end{itemize}
In the last case, one can always ensure that only one generator anti-commutes with $g$ by performing Gaussian elimination.
As long as $\rho_n, \rho_e$ are stabilizer states and $U_{SE}, U_{SA}$ are in the Clifford group, the noise and QE operations are also efficiently simulated. Specifically for noise, the partial trace operation on the environment qubits is performed by eliminating all generators with non-trivial support on the environment qubits after Gaussian elimination. 

To calculate the entanglement, it has been shown that the entanglement entropy of a subsystem $M$ is:
\begin{equation}
S_M = |M| - |G_M|
\label{eq:one}
\end{equation}
where $|M|$ is the number of qubits in $M$ and $|G_M|$ is the number of generators for the stabilizer group who only have non-trivial support in $M$~\cite{li2019}.This can also be practically calculated by tracing out $M^c$, with the number of remaining non-identity generators corresponding to $|G_M|$.

Although the above method is already efficient, in practice, we may have to store $\sim O(L^4)$ ancilla qubits to incorporate the QE operation. This still poses a significant challenge to memory and gate operation efficiency. However, we can exploit the fact that every ancilla qubit is acted upon non-trivially only once, and we are ultimately only interested in the conditional entanglement entropy. This numerical method, first proposed in~\cite{kelly2024a}, is succinctly reviewed here. Each time an ancilla qubit is coupled to the system, we can perform Gaussian elimination and then discard stabilizers that have non-trivial support only on the ancilla qubit and eliminate the remaining stabilizer’s support on the ancilla qubit. This approach prevents the number of generators from increasing linearly with time, ensuring the memory complexity to be $O(L^2)$. It’s worth noticing that the remaining generators may not commute with each other since we doesn’t keep track of the possibly non-trivial part on the ancilla qubits.

We denote the total number of generators discarded by $x$. Notice that these generators won’t be updated during the subsequent evolution, so they reamin being non-trivial only on the ancilla qubits. Finally, we would like to calculate $S(M|A) = S(M,A)-S(A)$. For $S(M,A)$, we can traceout $M^{c}$ and denote the number of stabilizers that are not identity as $y$. Using Eq.~(\ref{eq:one}), we have $S(M, A) = |M| + |A| - x - y$. On the other hand, $S(A)$ is simply $|A| - x$ since every stabilizer with non-trivial support only on the ancilla qubits are already discarded. It follows that:
\begin{equation}
S(M|A) = |M| - y.
\end{equation}
Remarkably, the dependence on the number of ancilla qubits and the number of discarded generators cancels out, allowing the conditional entanglement entropy to be calculated using only the generators of the final system state. 

\subsection{(2+1)-d simulation and data collapse}

\begin{figure}[t]
\begin{center}
\includegraphics[width=5in, clip=true]{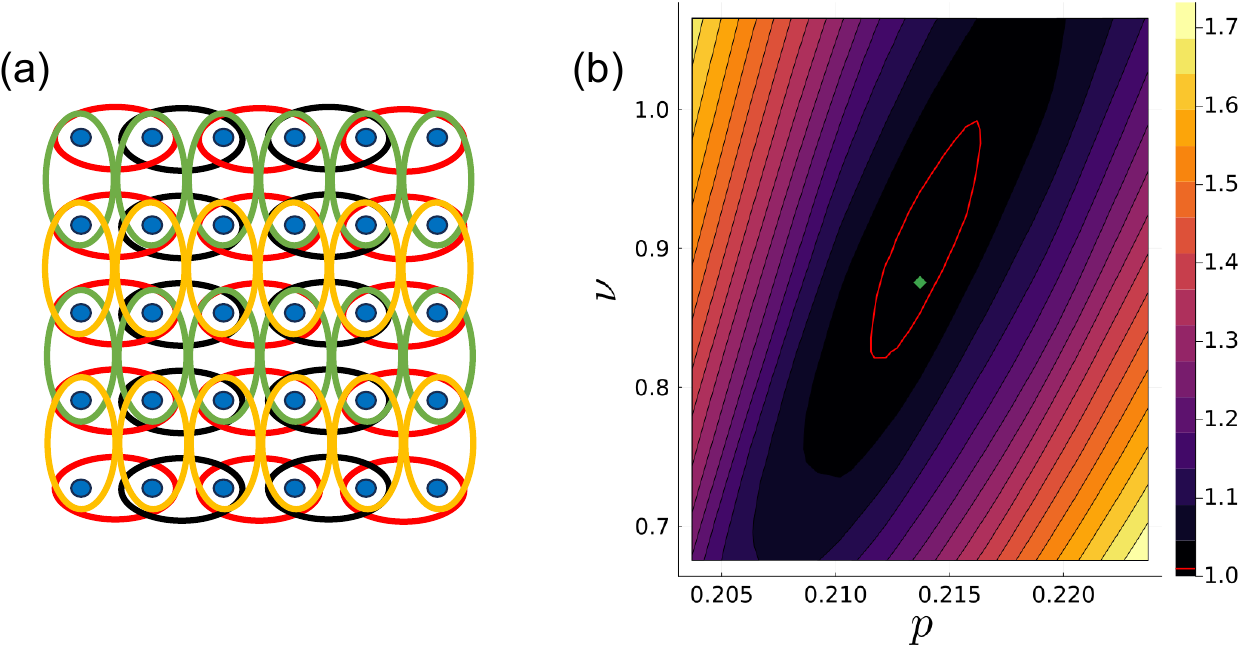}
\end{center}
\caption{(a) Brickwall structure of the $(2+1)$-d quantum circuit. Each time step consists of four unitary layers, composed of red, black, orange, and green gates, as indicated in the figure. The sequence of layers is red, yellow, black, and green.(b) Determine the uncertainty in $p_c$ and $\nu$. The residues are rescaled by dividing the $\epsilon^{\text{min}}$. Red circle is where the residue equals $1.01\epsilon^{\text{min}}$ and the green diamond is where the minimal point is. The uncertainty can be approximated to be $p_c=0.214\pm 0.002$ and $\nu=0.9 \pm 0.1$.}
\label{fig2}
\end{figure}
In the $(2+1)$-d setting, the brickwall structure of the unitary gates is designed as shown in Fig.~\ref{fig2}(a). Each time step consists of four unitary layers, applied in the order of red, yellow, black, and green gates. Measurement, noise and QE operation occur between every two layers. Each unitary layer consists of commuting unitary gates, with every qubit being acted upon by one gate per layer, subject to periodic boundary conditions. To mitigate the effect of periodicity on entanglement quantities, which depend on a specific cut across the system, each quantity at a given time step is first evaluated after each unitary layer and then averaged over the four layers. MIPT in the noiseless case happens at $p\sim 0.3$, which agrees with the result in~\cite{lunt2021}.

To accurately determine the critical point and get the critical exponent, we calculate the conditional tripartite mutual information $I_3$. Previous studies~\cite{sierant2022, lunt2021} have found that $I_3$ behaves as
\begin{eqnarray}
I_3 = \left\{\begin{matrix}\mathcal{O}(L^{2}), && p < p_c
 \\\mathcal{O}(1), && p = p_c
 \\0, && p > p_c
\end{matrix}\right.
\end{eqnarray}
in a $(2+1)$-d quantum circuit with the partition of the system as shown in the main text. We then collapse the $I_3$ data to the scaling form $I_{3} = f((p-p_c)L^{1/\nu})$. The data collapse procedure follows the method described in~\cite{qian2024} and proceeds as follows. For a given combination of $p_c$ and $\nu$, we rescale a particular data point $(p, L, I_3)$ as: 
\begin{equation}
    x = (p-p_c)L^{1/\nu},\ \ y=I_3.
\end{equation}
After rescaling all the data points, we fit the rescaled data with a $12$-th order polynomial and calculate the residue for the best fit. The residue $\epsilon(p_c, \nu)$ is then defined as the target function. By applying the Nelder-Mead algorithm, we find the minimal point $(p_c^{\text{min}}, \nu^{\text{min}})$ and the minimal residue $\epsilon^{\text{min}}$. To estimate the uncertainty in $p_c$ and $\nu$, we further plot the residue in the parameter space around the critical point. We set the threshold to be $1.01\epsilon^{\text{min}}$ to determine the uncertainty. Fig.~\ref{fig2}(b) shows an example for estimating the error for the dephasing case with $q=0.1$.

\begin{figure}[t]
\begin{center}
\includegraphics[width=5in, clip=true]{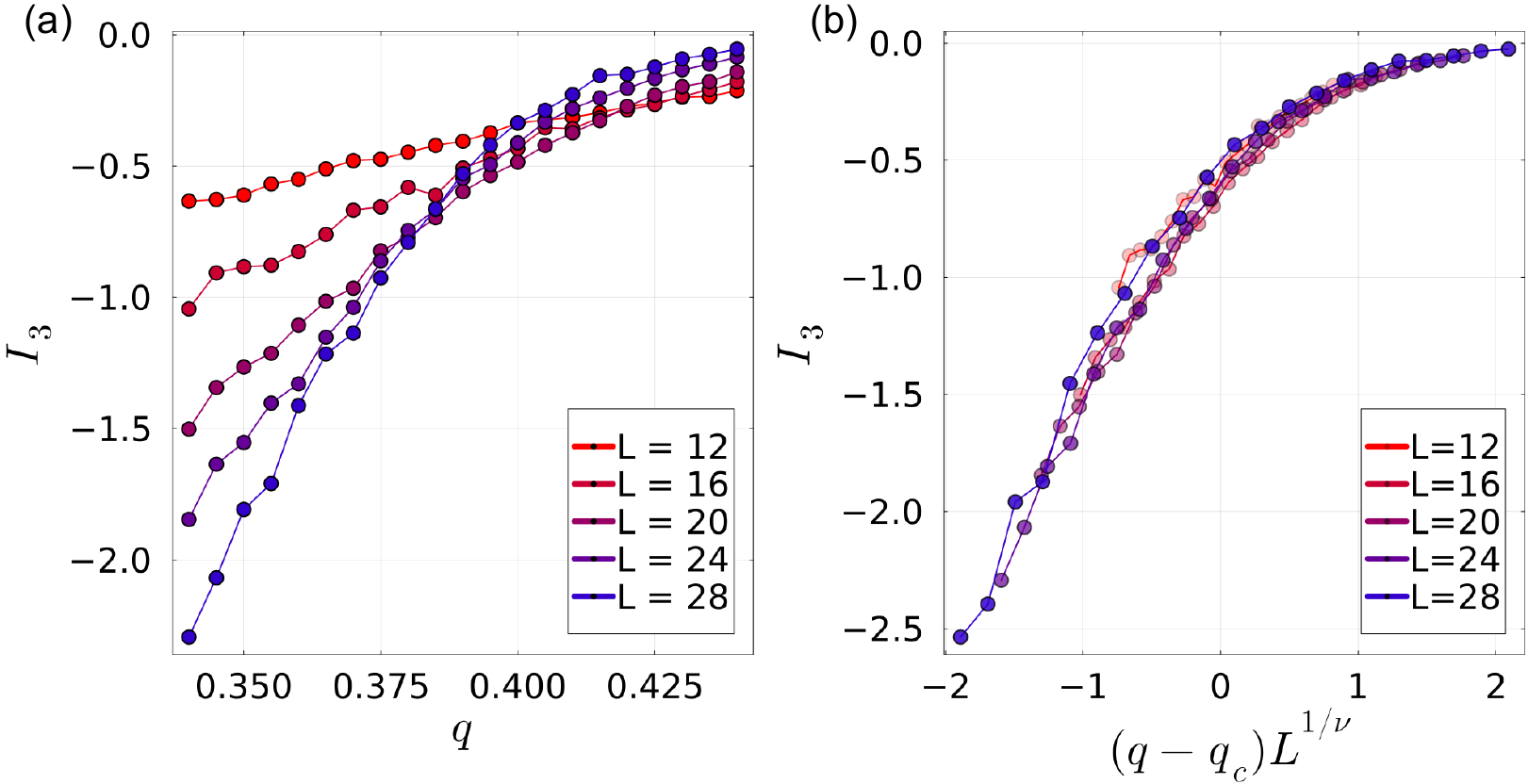}
\end{center}
\caption{Numerical results for dephasing noise for phase transition along the $p = 0$ line. Each data point is averaged over $5\times 10^{4}$ realizations. }
\label{fig3}
\end{figure}

We further investigate the phase transition along the $p = 0$ line by tuning $q$, corresponding to the disorder transition entirely at $T=0$. We focus on the dephasing noise and the symmetric QE operation, with the results presented in Fig.~\ref{fig3}. The critical point is identified at $q_c = 0.387 \pm 0.003$, with a critical exponent $\nu = 0.9 \pm 0.1$. For the $3$d random field Ising model, it has been shown that the finite temperature critical point flows to a zero-temperature fixed point under renormalization group~\cite{bray1985}. This is consistent with our  numerical findings, which indicate that the critical exponent are identical for the transitions at $q=0.1$ and $p=0.0$. Additional computational resources are required to conduct numerical simulations on larger system sizes with more samples, which will enable a more accurate determination of the universality class.

\begin{figure}[t]
\begin{center}
\includegraphics[width=5in, clip=true]{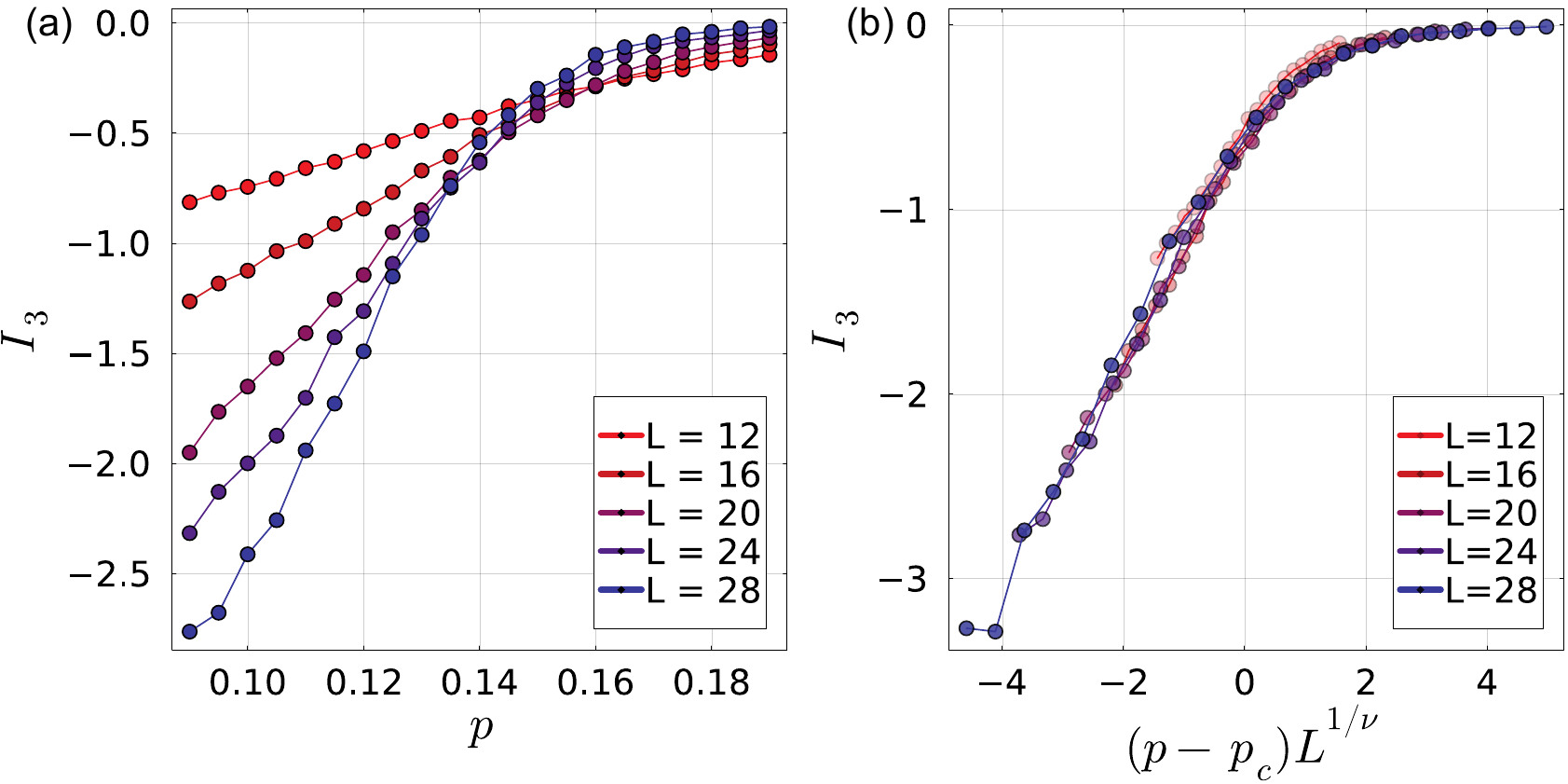}
\end{center}
\caption{Numerical results for dephasing noise for phase transition along the $q = 0.2$ line. Each data point is averaged over $5\times 10^{4}$ realizations. The critical point is determined to be at $p_c = 0.138(2)$, with $\nu = 0.73(7)$.}
\label{fig4}
\end{figure}

\begin{figure}[t]
\begin{center}
\includegraphics[width=5in, clip=true]{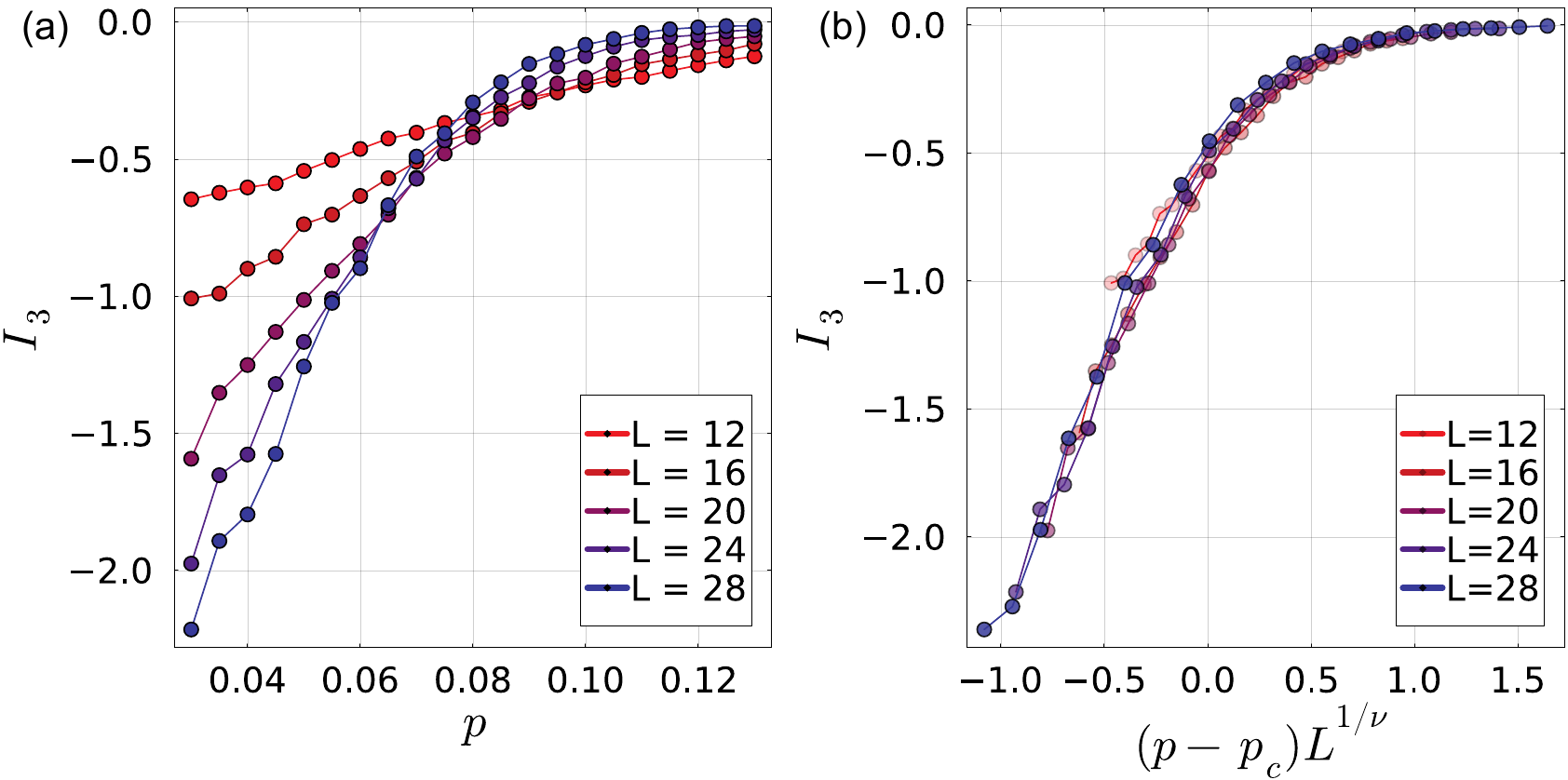}
\end{center}
\caption{Numerical results for dephasing noise for phase transition along the $q = 0.3$ line. Each data point is averaged over $5\times 10^{4}$ realizations. The critical point is determined to be at $p_c = 0.070(3)$, with $\nu = 1.0(1)$.}
\label{fig5}
\end{figure}

To investigate the behavior at higher noise rates, we simulate the cases with $q=0.2$ and $q=0.3$. The corresponding results are presented in Fig.~\ref{fig4} and Fig.~\ref{fig5}, respectively. Notably, the phase transition remains observable even at these relatively high noise rates, although the critical point shifts to lower values. Through data collapse, we determine the critical points to be $p_c = 0.138(2)$ and $p_c = 0.070(3)$, for $q=0.2$ and $q=0.3$, respectively. 

\subsection{Purification dynamics}

\begin{figure}[t]
\begin{center}
\includegraphics[width=7in, clip=true]{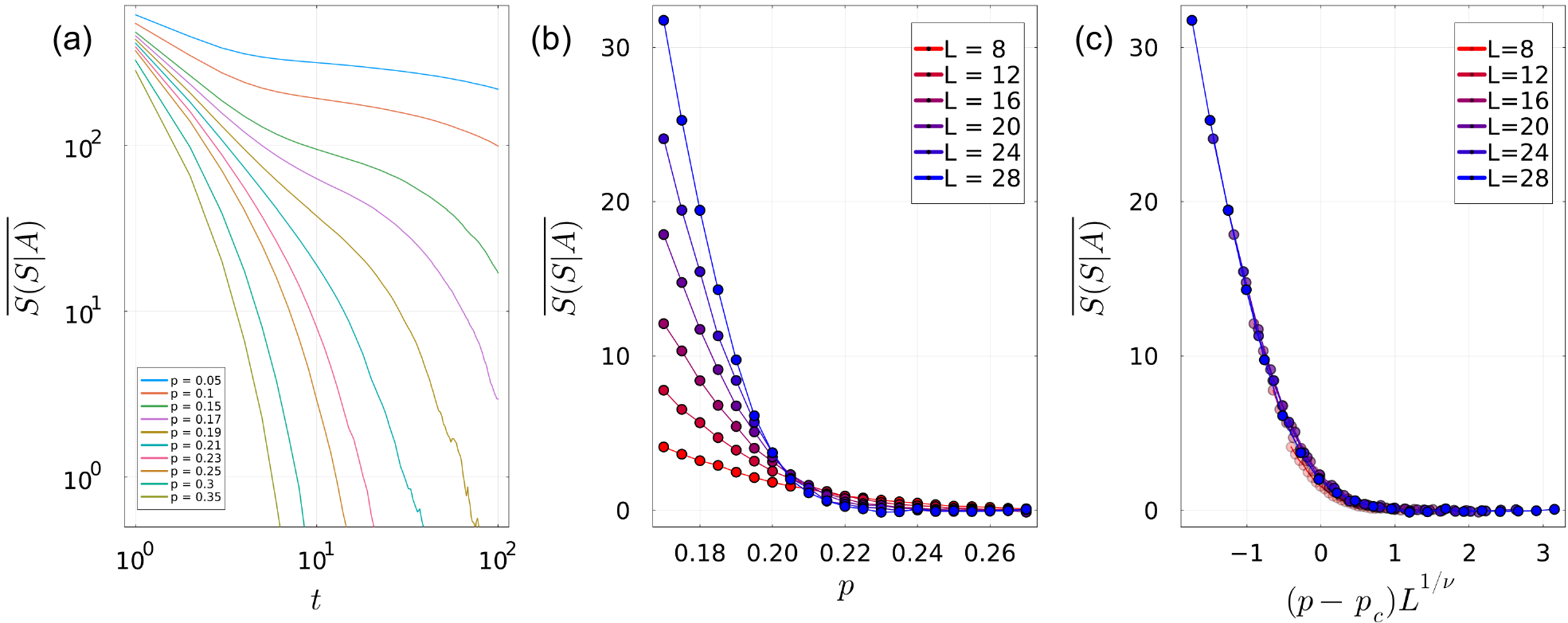}
\end{center}
\caption{Numerical simulation for purification dynamics. The input state is taken to be the maximally mixed state. Every data point is averaged over $5 \times 10^{4}$ realizations. (a) $\overline{S(S|A)}$ for dephasing noise and symmetric QE operation, with $q=0.1$. The system size is taken to be $L = 28$. (b-c) $\overline{S(S|A)}$ at $t=L$ for different system sizes. The state is purified with a zero $\overline{S(S|A)}$ for $p>p_{c}$, while $\overline{S(S|A)}$ is extensive for $p<p_{c}$. The data collapse clearly shows a phase transition.}
\label{fig6}
\end{figure}

The MIPT has been demonstrated to coincide with a dynamical purification transition~\cite{gullans2020a}. This phenomenon is investigated by initializing the system with a maximally mixed state, $\rho_{S} = \frac{1}{2^{N}}\mathbb{I}$, and observing the time required for state purification. Conceptually, this maximally mixed state on subsystem $S$ can be interpreted as a component of a maximally entangled pure state spanning $S$ and a reference system $S'$. The entanglement between $S$ and $S'$ persists as long as the state of $S$ remains mixed.
The system exhibits distinct behaviors depending on the measurement rate $p$. For $p < p_c$, the purification time scales exponentially with system size. This behavior effectively constitutes a dynamical quantum error correction code. For $p > p_c$, the state undergoes rapid purification, with the purification time scaling polynomially with system size. Furthermore, this scenario can be analyzed from a quantum information theoretic perspective. If we consider the circuit as a quantum channel $\mathcal{N}$ that maps an input state on $S'$ to an output state on $S$, the averaged entanglement entropy of the final state corresponds to the coherent information $I_{C}(S'>S)$ for this channel. Extending this framework to include ancillary qubits $A$, the channel now maps the input state to an output state distributed across both $S$ and $A$. In this extended scenario, our focus shifts to the coherent information $I_{C}(S'>S\cup A)$, which is equivalent to the conditional entanglement entropy $\overline{S(S|A)}$~\cite{kelly2024a}.

In Fig.~\ref{fig6}, we show numerical results for the purification dynamics in the presence of  dephasing noise and the symmetric QE operation, at a rate $q=0.1$. $\overline{S(S|A)}$ remains to be finite over a timescale polynomially in system size when $p < p_c$. Conversely, $\overline{S(S|A)}$ rapidly decays to zero for $p>p_c$. Focusing on $t=L$, we scan through different system sizes and collapse the data according to the form $\overline{S(S|A)} = g((p-p_c)L^{1/\nu})$. A clear crossing is observed and the critical point is determined to be at $p_c = 0.206(1)$, with the critical exponent $\nu = 0.86(5)$. Notably, these values are close to that of the entanglement phase transition, as shown in the main text. Thus, we argue that these two transitions represent two facets of the same underlying transition, which is the same to the case for the MIPT without noise.

\subsection{Other noises}

\begin{figure}[t]
\begin{center}
\includegraphics[width=7in, clip=true]{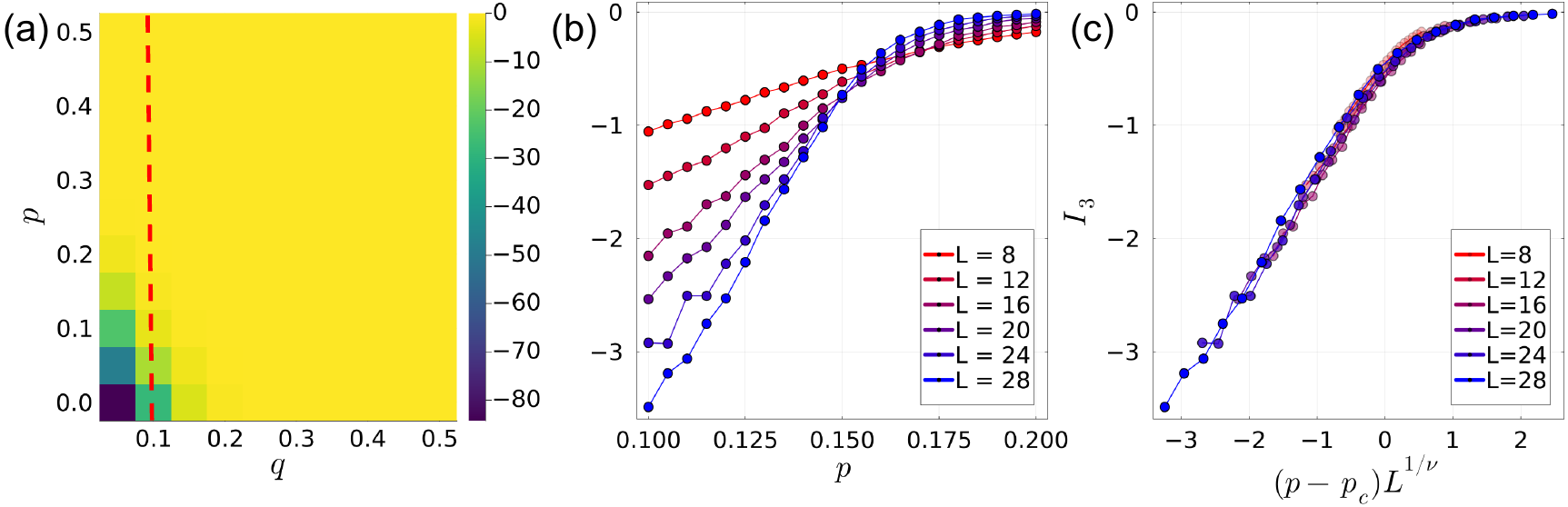}
\end{center}
\caption{Resetting noise. (a) $I_3$ vs $(p,q)$ with $L=20$. The red line $q=0.1$ is further examined in (b), (c), which demonstrate MIPT in the presence of noise and QE operation. Each data point is averaged over $5\times10^4$ realizations.}
\label{fig7}
\end{figure}

\begin{figure}[t]
\begin{center}
\includegraphics[width=7in, clip=true]{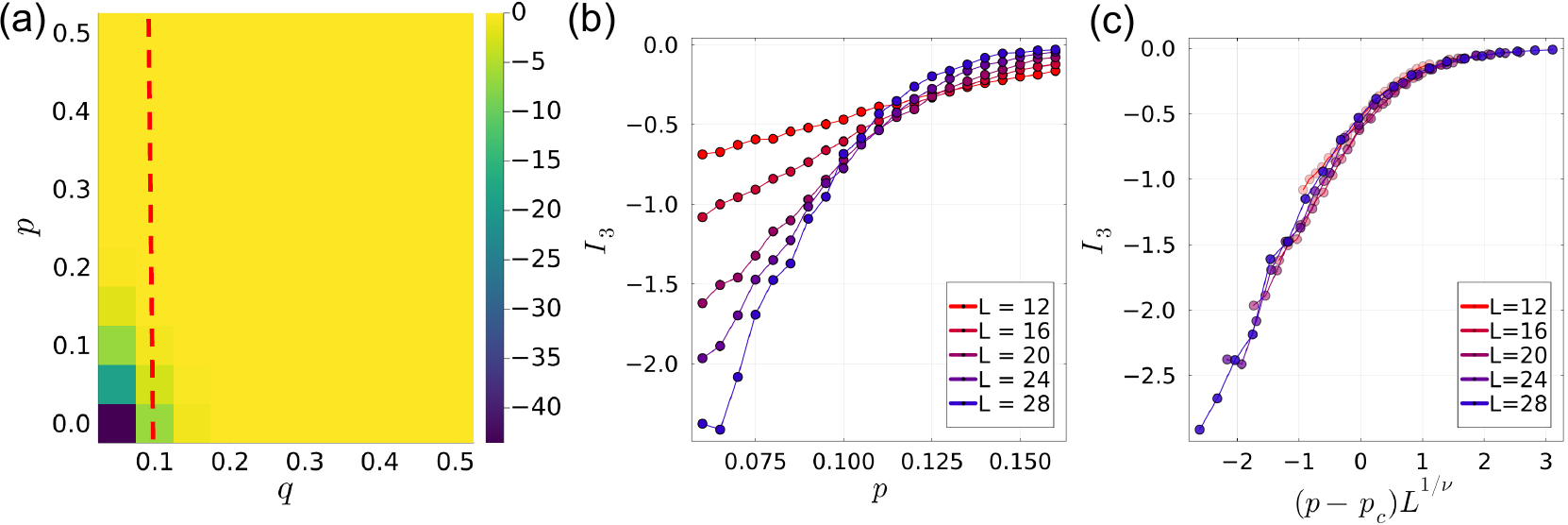}
\end{center}
\caption{Depolarizing noise. (a) $I_3$ vs $(p,q)$ with $L=20$. The red line $q=0.1$ is further examined in (b), (c), which demonstrate MIPT in the presence of noise and QE operation. Each data point is averaged over $5\times10^4$ realizations.}
\label{fig8}
\end{figure}

The approach of protecting the MIPT from noise through quantum-enhanced operations is general and not limited to a specific type of noise. In the main text, we presented numerical results for dephasing noise and symmetric QE operations. Here, we further demonstrate that MIPT can be protected against resetting and depolarizing noise in a similar manner. The absence of MIPT in the presence of resetting and depolarizing noise has been discussed in Ref.~\cite{liu2023e} and Ref.~\cite{weinstein2022}, respectively. The numerical results for resetting noise and depolarizing noise, along with the corresponding symmetric QE operations, are shown in Fig.~\ref{fig7} and Fig.~\ref{fig8}, respectively. The critical points are $p_c^{\text{reset}}=0.157(3)$ and $p_c^{\text{depo}}=0.106(3)$, with the critical exponents determined by data collapse being $\nu^{\text{reset}}=0.8(1)$ and $\nu^{\text{depo}}=0.8(1)$. Due to the similarity in critical exponents for different types of noise, we conjecture that they belong to the same universality class. 
Additionally, it is noteworthy that the volume law phase region is smallest for depolarizing noise and largest for dephasing noise, consistent with the order of random field strengths inferred from Eq.~(\ref{eq:S11}), Eq.~(\ref{eq:S12}) and Eq.~(\ref{eq:S17}).

\begin{figure}[t]
\begin{center}
\includegraphics[width=5in, clip=true]{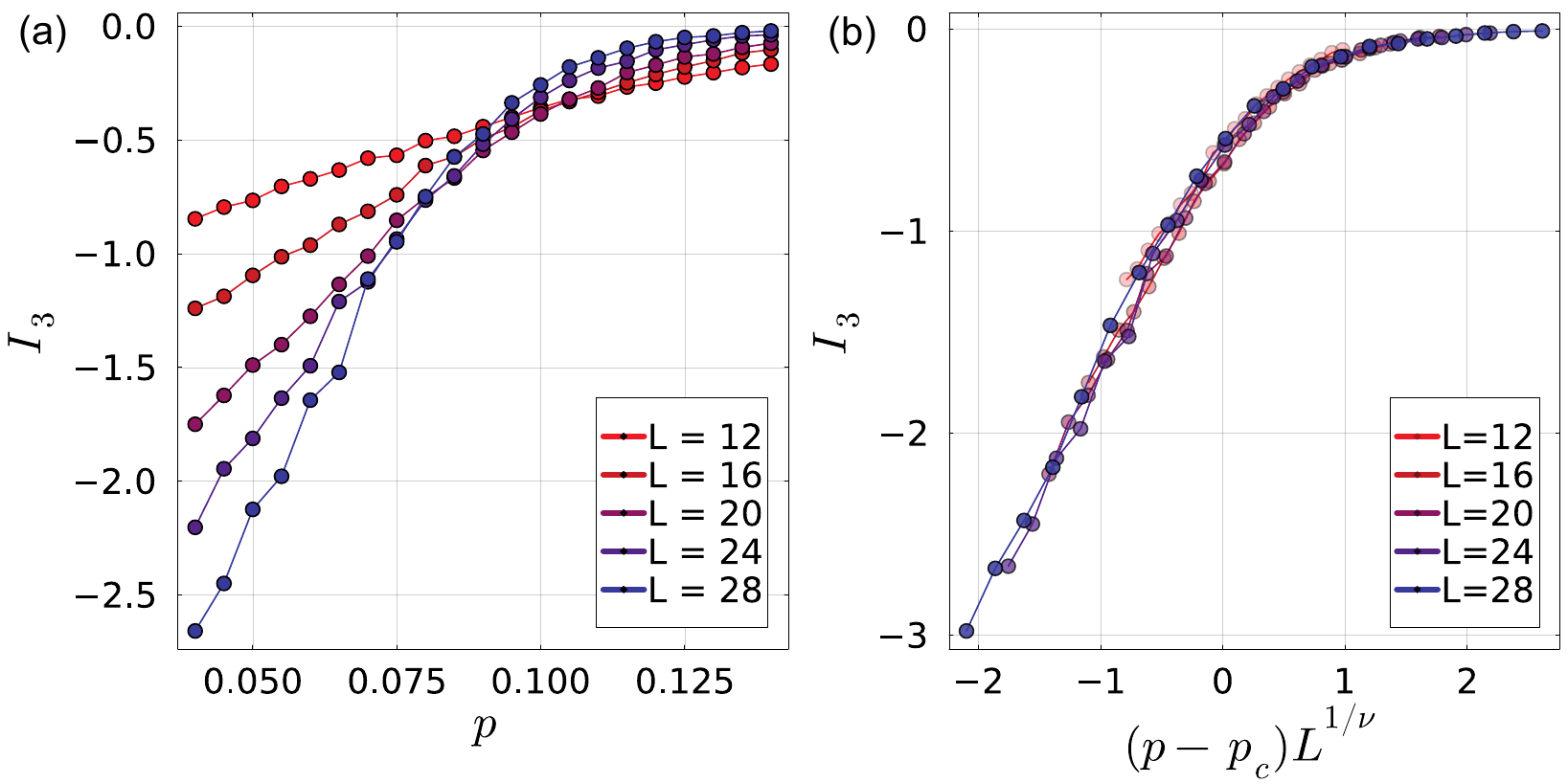}
\end{center}
\caption{Multiple noises. (a) $q^{\text{deph}}=0.1$ and $q^{\text{reset}}=0.1$. The critical point is located by data collaping in (b), with $p_c = 0.085(3)$ and $\nu=0.86(8)$. Each data point is averaged over $5\times10^4$ realizations.}
\label{fig9}
\end{figure}

To further demonstrate the effectiveness of QE operations in protecting the MIPT, we consider a scenario with multiple kinds of noise in the circuit. For each type of noise, a corresponding symmetric QE operation is applied. Specifically, we analyze the case where both dephasing noise and resetting noise are present, each with a rate of $q^{deph} = q^{reset} = 0.1$, resulting in a total noise rate of 10\%. The result, shown in Fig.~\ref{fig9}, reveal a clear phase transition occurring at a critical measurement rate of $p_c=0.085(3)$. 

\subsection{Determine noise rate}
\begin{figure}[t]
\begin{center}
\includegraphics[width=7in, clip=true]{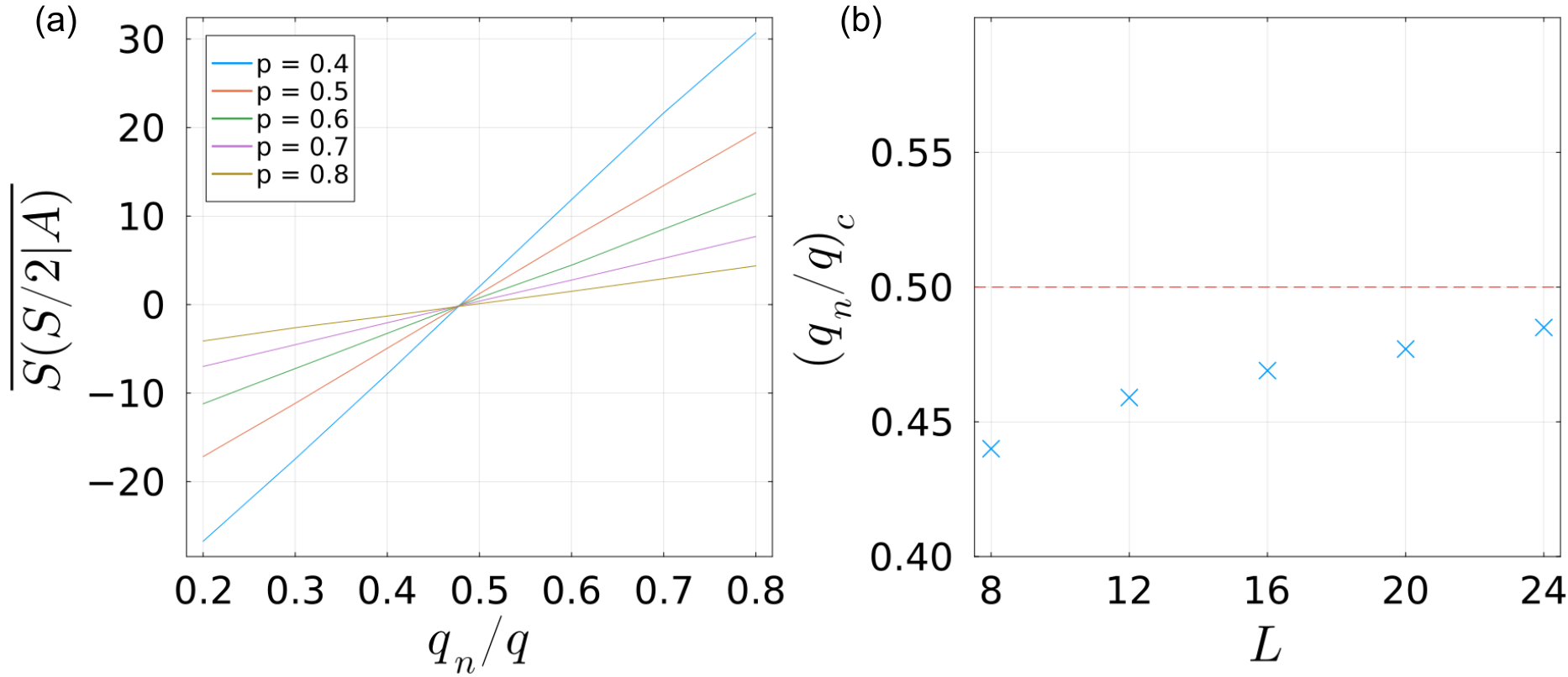}
\end{center}
\caption{Method of determining the noise rate. (a) Numerical results for $L=20$. We fix the total rate $q=0.2$ and vary the ratio $q_{n}/q$. Each data point represents an average over 2,000 realizations. (b) The intersection point $(q_{n}/q)_c$ as a function of different system sizes $L$. The line $(q_{n}/q)_c = 0.5$ corresponds to the thermodynamic limit.}
\label{fig10}
\end{figure}

The requisite aAEE symmetry for observing the MIPT can be leveraged to estimate the error rate in the system. In the thermodynamic limit, the conditional bipartite entanglement entropy $S(S/2|A)$ exhibits invariance upon the system's transition into the area law regime, contingent on the preservation of aAEE symmetry. In the absence of this symmetry, $S(S/2|A)$ manifests distinct behaviors dependent on the relative magnitudes of the $q_n$ and $q_e$. When $q_n > q_e$, $S(S/2|A)$ manifests a monotonic decrease from a positive value as the measurement rate increases. Conversely, when $q_n < q_e$, $S(S/2|A)$ exhibits a monotonic increase from a negative value. Exploiting this phenomenon, we propose a method for determining the error rate of the system. By conducting a parametric scan across various $q_e$ values while maintaining high measurement rates—thus ensuring the system's residence in the area law phase—one can discern an intersection point among curves corresponding to different measurement rates. This point, characterized by the independence of $S(S/2|A)$ with respect to the measurement rate $p$, precisely corresponds to the condition $q_e = q_n$. We conducted extensive numerical simulations on finite-size systems to explicitly demonstrate this method. The results for a system of $L=20$ are presented in Fig.~\ref{fig10}(a). To investigate the scaling behavior and approach to the thermodynamic limit, we varied the system size. As illustrated in Fig.~\ref{fig10}(b), there is a clear trend that as the system dimensions increase, the intersection point of the curves converges asymptotically towards to $q_n=q_e$. It is worth noting that this method is effective when a single dominant noise source is present, which is typically the case in experiments. However, if multiple noise sources are involved, determining the noise rate for each becomes challenging. Therefore, further improvements or the development of new methods for estimating noise rates would be highly valuable for future work.

\subsection{Unequal rates}
\begin{figure}[t]
\begin{center}
\includegraphics[width=5in, clip=true]{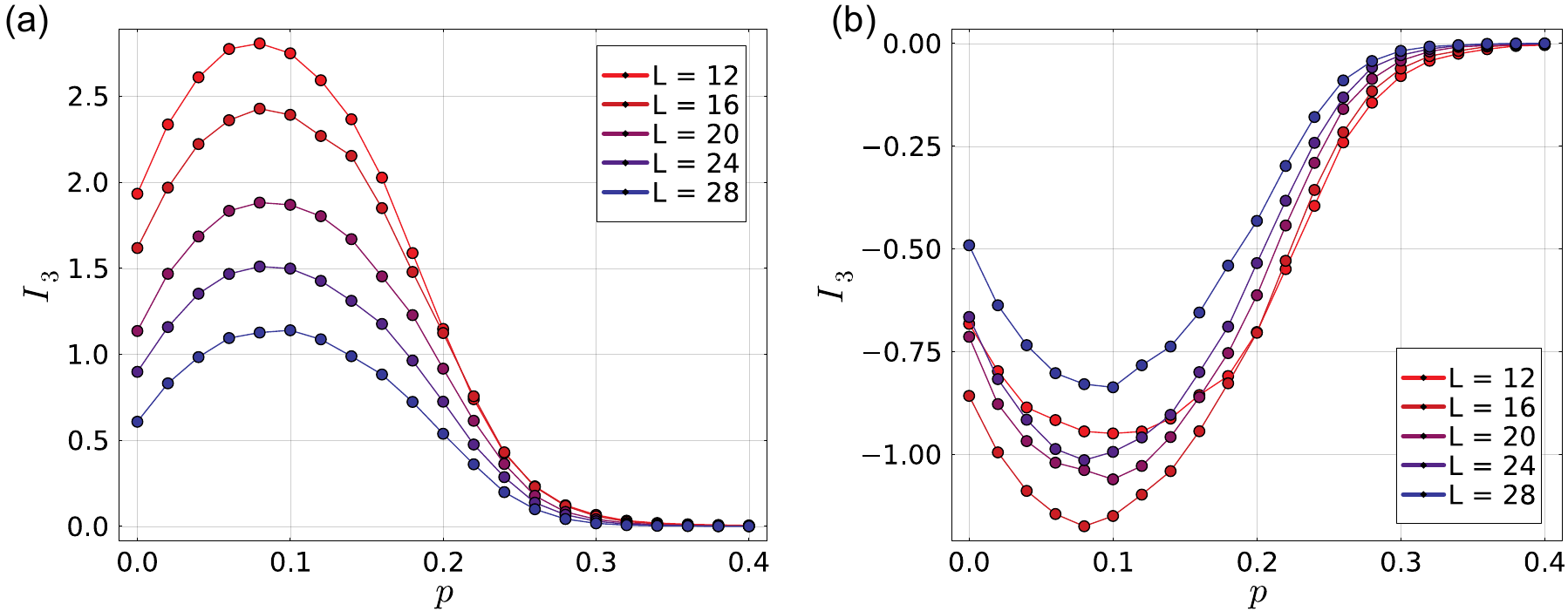}
\end{center}
\caption{Absence of phase transition with unequal $q_n$ and $q_e$. We consider dephasing noise with the corresponding symmetric QE operation. Every data point is averaged over $10^4$ realizations. (a) $q_n = 0.04$ and $q_e = 0.06$. (b) $q_n = 0.06$ and $q_e = 0.04$.}
\label{fig11}
\end{figure}

In the main text, we provided an analytical explanation for the necessity of the zero-field condition in observing the MIPT. When a non-zero net field is present, the resulting CEE consistently follows a volume-law scaling, thereby preventing the phase transition. To further substantiate this argument, we performed numerical simulations by examining scenarios where the rate of dephasing noise differs from that of the corresponding symmetric QE operation. Specifically, we consider two cases: $q_n=0.04$ with $q_e=0.06$, and $q_n=0.06$ with $q_e=0.04$. In both cases, the zero-field condition is violated, and as a result, the aAEE symmetry is no longer preserved. Consequently, in the calculation of the tripartite mutual information $I_3$, we have $S’(\text{abc}) \propto \text{Volume}(\text{abc}) \neq S’(\text{a})$. Given that the CEE exhibits volume-law scaling, we expect $I_3$ to be zero for all measurement rates $p$. The results are shown in Fig.~\ref{fig11}. While finite-size effects are present, it is clear that $I_3$ approaches zero in the thermodynamic limit and does not exhibit crossings for sufficiently large system sizes.

